\documentclass[12pt,preprint]{aastex}
\usepackage{graphicx}
\usepackage{txfonts}
\usepackage{natbib}

\slugcomment{Submitted  to  AJ}

\shorttitle{Globular  clusters  in  M31}
\shortauthors{Rich  et  al.\ }

\begin{document}


\title{HST/WFPC2  Color-magnitude  Diagrams  for  Globular  Clusters  in
M31\footnote
{Based  on  observations  made  with  the  NASA/ESA  Hubble  Space 
Telescope  at  the  Space  Telescope  Science  Institute.  STScI  is  operated  
by  the  Association  of  Universities  for  Research  in  Astronomy,  Inc.
under  NASA  contract  NAS  5-26555.}}


\author{R.M.  Rich\altaffilmark{}}    
\affil{Dept.  of  Phys.  \&  Astron.,  UCLA,  Los  Angeles,  CA  90095-1562,  USA}
\email{rmr@astro.ucla.edu}

\author{C.E.  Corsi\altaffilmark{}}    
\affil{Osservatorio  Astronomico,  V.le  Parco  Mellini  84,  00136  Roma,  Italy}
\email{corsi@mporzio.astro.it}

\author{C.  Cacciari,  L.  Federici,  F.  Fusi  Pecci  \altaffilmark{}}
\affil{Osservatorio  Astronomico,  Via  Ranzani  1,  40127  Bologna,  Italy}
\email{cacciari@bo.astro.it, luciana@bo.astro.it, flavio@bo.astro.it}

\and

\author{S.G.  Djorgovski\altaffilmark{}}
\affil{Palomar  Observatory,  CalTech,  USA}
\email{george@astro.caltech.edu}






\begin{abstract}
We  report  new  photometry  for  10  globular  clusters  in  M31,
observed  to  a  uniform  depth  of  four  orbits  in  F5555W(V)  and
F814W(I)  using  WFPC2  on  board  HST.      Additionally  we  have
reanalyzed  HST  archival  data  of  comparable  quality,  for  2  more
clusters.    A  special  feature  of  our  analysis  is  the  extraordinary  care
taken  to  account  for  the  effects  of  blended  stellar  images  and
required  subtraction  of  contamination  from  the  field  stellar  populations
in  M31  in  which  the  clusters  are  embedded.    We  thus  reach  1  mag  fainter
than  the  horizontal  branch  (HB)  even  in  unfavorable  cases.    We  also
show  that  an  apparent  peculiar  steep  slope  of  the  HB  for
those  clusters  with  blue  HB  stars  is  actually  due  to  blends  between
blue  HB  stars  and  red  giants.

We  present  the  color-magnitude  diagrams  (CMDs)  
and  discuss  their  main  features  also  in  comparison  
with  the  properties  of  the  Galactic  globular  clusters.  
This  analysis  is  augmented  with  CMDs  previously  
obtained  and  discussed  by  Fusi  Pecci  et  al.\  (1996)  on  8  other  M31  clusters.  
We  report  the  following  significant  results:

1.  The  locus  of  the  red  giant  branches  give  reliable  photometric  metallicity  
determinations  which  compare  generally  very  well  with  ground-based  
integrated  spectroscopic  and  photometric  measures,  as  well  as  
giving  good  reddening  estimates.    

2.  The  HB  morphologies  follow  the  same  behavior  with  metallicity  
as  the  Galactic  globular  clusters,  with  indications  that  the  2nd-parameter
effect  can  be  present  in  some  clusters  of  our  sample.    However,  at  [Fe/H]=$\sim  -1.7$
we  observe  a  number  of  clusters  with  red  HB  morphology  such  that
the  HB  type  versus  [Fe/H]  relationship  is  offset  from  the  Milky  Way  and  resembles
that  of  the  Fornax  dwarf  spheroidal  galaxy.      One  explanation  for  the  offset
is  that  the  most  metal  poor  M31  globular  clusters  are  younger  than  their
Milky  Way  counterparts  by  1-2  Gyr;  further  study  is  required.

3.    The  M$_V$(HB)  versus  [Fe/H]  relationship  has  been  re-determined  and  the  
slope  ($\sim$0.20)  is  very  similar  to  the  values  derived  from  RR  Lyrae  stars  
in  the  MW  and  the  LMC.
The  zero-point  of  this  relation  (M$_V$  =  0.51  at  [Fe/H]=--1.5)  is  based  on    
the  assumed  distance  modulus  (m--M)$_0$(M31)=24.47$\pm$0.03,  
and  is  consistent  with  the  distance  scale  that  places  the  LMC  at  
(m--M)$_0$(LMC)=18.55.  
    

\end{abstract}


\keywords{clusters:  globular;  photometry:  CMDs;  individual  Messier  number:  M31;
stellar  populations}


\clearpage  

\section{Introduction}

Among  the  Local  Group  galaxies,  M31  has  the  largest  population  of  
globular  clusters  (GC)  (460$\pm$70,  see  Barmby  \&  Huchra  2001)  and
is  the  nearest  analog  of  the  Milky  Way.
Its  distance  from  the  Milky  Way  (MW),  $\sim$780 kpc, is  large  enough  
so  that  the  dispersion  in  distance  
modulus  of  the  GC  system  can  be  considered  to  be  small  (50  kpc  
correspond  to  $\delta(M-m)\sim$  0.15  mag), and  
hence  the  globulars  are  nearly  at  the  same  distance  to  us.  Also,  
their  almost  stellar  appearance  (10  pc  correspond  to  $\sim$ 2.6  arcsec)  
allows  an  easy  study  of  their  integrated  properties  from  the  ground.    

On  the  other  hand,    M31  is  also  close  enough  that  individual  stars  in  
GCs    can  be  resolved  and  measured  with  HST  (and  with  very  large  ground-based  
telescopes  equipped  with  powerful  adaptive  optics  systems.
Therefore,  good  Color-Magnitude  diagrams
(CMDs)  can  be  obtained,  reaching  well  below  the  Horizontal  Branch  (HB),    
as  was  shown  by  the  early  HST  surveys  (Ajhar  et  al.\  1996;  Rich  et  al.\  1996;
Fusi  Pecci  et  al.\  1996  --  hereafter  Paper  I, Holland et al.\ 
1997; Jablonka et al.\ 2000; Stephens et al.\ 2001),  and  further  confirmed  
by  recent  very  deep  observations  using  the  Advanced  Camera  for
Surveys  on  board  HST  (Brown  et  al.\  2003,  2004a,b).  In  contrast to adaptive  
optics, HST not only avoids the vagaries  of  a spatially and temporally  variable  point  
spread  function, but also gives imagery in the optical bandpasses
which are most sensitive to metal line blanketing and have
a vast heritage of prior studies.

While  in  many  ways  M31  is  similar  to  the  Milky  Way,  there  are  important
differences.    Brown  et  al.\  find  evidence  for  an  age  dispersion  in  the
halo,  with  a  metal  rich  population  as  young  as  6-8  Gyr  old.    The  halo
itself  appears  to  be  dominated  by  stars  acquired  from  the  ingestion  of
other  stellar  systems,  the  signature  of  which  is  a  halo  of  complex  and
irregular  morphology,  including  a  giant  tidal  stream  (Ferguson  et  al.\  2002).
Finally,  there  has  been  a  long  standing  question  of  chemical  peculiarities
in  the  M31  clusters,  likely  an  enhancement  of  nitrogen  (see Burstein
et  al.\  2004,  and  Rich  2004).  

The  above  mentioned  issues  make  the  detailed  observations  of
the  M31  GC  system  extremely  valuable  for  very  important  comparisons
with    the  analog  systems  in  the  MW  and  in  external  galaxies.  The  M31  GCs  
may  be  used  as  templates  and  a  sort  of  bridge  between  fully  resolved  systems  
and  totally  unresolved  ones  in  the  study  of  stellar  populations,    
with  important  implications  for  galaxy  formation  theories  and  cosmology.

From  the  observational  evidence  collected  so  far  (see Perrett et al.\ 2002; 
Barmby 2003; Galleti et al.\ 2004, and  references therein),  
the  M31  GCs  show  all  indications  
of  being  very  similar  to  the  MW  globulars.  
Although  no  ``direct''  estimate  of  age,  nor  accurate  metal  abundances  of  
individual  star  members,  are  available  for  any  M31  GC  except  one  (G312,  
see  Brown  et  al.\  2004b),  most  of  them  are  presumably  as  old  as,  and  slightly  
more  metal-rich  than    the  MW  globulars.  
The  two  GC  systems  occupy  similar  loci  in  the  ``fundamental  plane''  
(McLaughlin  2000)  and  seem  to  have  similar  M/L,  structural  parameters  
(Fusi  Pecci  et  al.\  1994;  Djorgovski  et  al.\  1997;  Barmby  et  al.\  2002;  
Djorgovski  et  al.\  2003),
and  high  incidence  of  strong  X-ray  sources  (van  Speybroeck  et
al.\  1979;  Bellazzini  et  al.\  1995;  Di Stefano  et  al.\  2002,  
Trudolyubov \& Priedhorsky  2004).

However,  as  previously  mentioned,  the  globular  cluster  system  exhibits
noteworthy  contrasts  with  that  of  the  Milky  Way:

a)  By  comparing  integrated  GC  colors  with  stellar
population  models,  Barmby  et  al.\  (2001)  concluded  that
the  metal-rich  GCs  in  M31  are  younger  (by  4-8  Gyr)  than  the
metal-poor  ones.  A  similar  interpretation  was  suggested
to  explain  the  stronger  H$\beta$  lines  observed  in
metal-rich  M31  GCs  compared  to  Galactic  globulars  (Burstein
et  al.\  1984),  but    Peterson  et  al.\  (2003)  proposed  an  alternative
explanation  based  on  the  presence  of  old  blue  HB  stars
(see  also  Fusi  Pecci  et  al.\  2004).  

b)  The  metallicity  distribution  of  M31  GCs  is  clearly  bimodal
(Barmby  et  al.\  2000)  and  there  are  indications  (Huchra  et  al.
1991;  Perrett  et  al.\  2002)  for  a  systematic  difference  in  
kinematics  and  spatial  distribution  between  the  two  metallicity
groups,  somehow  supporting  possible  differences  in  the  formation
process  and  age  (Ashman \& Zepf  1998;  Saito \& Iye  2000).
In  particular,  Morrison  et  al.\  (2004)  claim  that  there  
is  a  subsystem  of  GCs  in  M31  with  thin  disk  kinematics,
whereas  no  GC  is  known  to  be  associated  with  the  Galactic  
thin  disk.

c)  There  are  indications  for  possible  variations  of  the  GC
Luminosity  Function  (Barmby  et  al.\  2001)  and  average  structural  
parameters    (Djorgovski  et  al.\  1997,  2003;  van  den  Bergh  2000;  Barmby  
et  al.\  2002)  with  galactocentric  distance  and  metallicity,  that  might    
be  ascribed  to  differences  in  age,  destruction/survival/capture  rate,  etc.

d)  There  is  clear  evidence  for  the  existence  of  streams
and  overdensities  associated  with  metallicity  variations
across  the  whole  body  of  M31  (Ibata  et  al.\  2001;  Ferguson
et  al.\  2002),  possibly  related  to  interactions  with  close
companions  (Bellazzini  et  al.\  2001;  Bekki  et  al.\  2002;
Choi  et  al.\  2002).  On  analogy  to  what  has  occurred  
between  the  Sagittarius  dSph  and  the  MW  (Ibata  et  al.\  1994),  it  may
be  conceivable  that  also  a  fraction  of  the  M31  GCs  was  
captured  and  differs  in  some  property  (age?  chemical  composition?  
kinematics?)  from  the  main  body  of  ``native''  clusters.  
In  this  respect  it  may  be
worth  noting  that  M32  does  not  appear  to  have  any  (residual)
GC  system  (van  den  Bergh  2003).


The  above  issues  can  be  studied  using  different  and
complementary  approaches.  For  example,  one  could
observe  the  integrated  light  of  GCs  in  specific  bands  or
indexes  (far  UV,  H$\beta$,  IR,  etc.)  sensitive  to
age,  or  metallicity,  or  peculiar  HB-morphology  (or,  more
probably,  a  hard-to-disentagle  mixture  of  them).
Or,  one  could  study  some  very  special  stellar
populations,  e.g.  the  variables  (see  the  detection  of  several  
possible  RR  Lyrae  candidates  in  four  M31  GCs  by  Clementini  et  al.\  2001),  
or  the  X-ray  sources.  However,  the  past  forty  years    of
research  on  Galactic  GCs  taught  us  that  one  can  hardly  hope  to  reach  
any  firm  and  unambiguous  conclusion  whithout  having  the  CMDs  and
(possibly)  the  spectra  of  individual  stars.

If  we  aim  to  assay  the  age  of  the  cluster  system,  it  is  possible  
to  do  so directly  by  the  measurement  of  main  sequence  photometry.
In  M31,  a  12  Gyr  old  population  has  a  main  sequence  turnoff  (TO)
at  V$\ge  $28.5  and  photometry  must  reach  1-2  mag  fainter  for
a  precise  age  measurement  for  the  oldest  stars.    This  has  been
accomplished  in  one  field  in  M31,  by  investing  120  orbits  of
imaging  with  HST+ACS  (Brown et al.\  2003,  2004a).  Due  to  extreme
crowding,  a  somewhat  less  stringent  age  constraint  was  determined
for  the cluster G312  (Brown  et  al.\  2004b)  in  the  same  deep  ACS  field.    
For the  foreseeable  future,  such  a  campaign  will  be  practical  for
a  very  small  number  of  fields.    Combining  spectroscopy  and  SEDs
offers  (in  principle)  another  approach  to  constraining  the  age.    This
approach  is  enjoying  an  increasing  level  of  success.    However,  
the  measurement  of  CMDs  to  below  the  horizontal  branch  gives
an  additional  age  constraint  and  a  powerful  means  of  comparing  the
M31  clusters  with  other  cluster  populations.      Knowledge  of  the  actual
CMDs  also  improves  the  accuracy  of  spectroscopic  and  SED  based
methods.

As  mentioned  above,  all  we  presently  know  about  the  GCs  in  M31  rests  
on  colors  and  metallicities  from  integrated  ground-based photometric  and
spectroscopic  observations (Barmby \& Huchra 2001; Perrett et al.\ 2002; Galleti et al.\ 
2004, and  references  therein),    
and  on  the  CMDs  of  the  few    clusters  previously  observed  with  HST  
(Ajhar et al.\ 1996; Rich et al.\ 1996; Fusi Pecci et al.\ 1994, 1996; Holland et al.\ 
1997; Jablonka et al.\ 2000; Stephens et al.\ 2001). 

With  the  HST-WFPC2  observations  (GO  program  6671,  P.I.  Rich)  we  present  here    
we  have  more  than  doubled  the  number  of  clusters  observed  with  HST.  
Preliminary  results  obtained  from  these  data  were  presented  by  Corsi  
et  al.\  (2000)  and  Rich  et  al.\  (2001).  
The  present  paper  reports  the  final  results  on  these  10  additional  
GCs  in  M31.    
In  the  following  discussion  we  add  two  more  
clusters    using  archive  WFPC2  data  that  were  obtained  
in  very  similar  conditions  (GO  program  5906,  P.I.  Holland).  
            
In  Sect.  2  we  present  a  description  of  the  data,  and  the  
data  reduction,    field  subtraction  and  calibration  procedures.  
The  presentation  and  analysis  of  the  results,    in  
Sect.  3,  include  some  discussion  on  the  overall  properties
of  the  CMDs  we  have  obtained  (considering  also  the  
CMDs  of  the  8  M31  GCs  described  in  Paper  I)  and  a  
general  comparison  with  typical  Galactic  globulars.  In  particular,  
we    discuss  the  Red  Giant  Branch  (RGB)  and  Horizontal  Branch  (HB)  
location  and  morphology,  and  the  estimate  of  parameters  such  
as  (photometric)  metallicity,  reddening,  HB-type,  HB-luminosity  level,  
and  M31  distance.  
A  summary  and  conclusions  can  be  found  in  Sect.  4.    

\section{The  data}

\subsection{Observations}

Our  target  GCs  were  selected  from  the  brightest  
(i.e.  most  populous)  objects,  over  a  wide  range  
of  radial  distances  and  metallicities,  taking  into  account  also  
the    clusters  that  had  been  observed  before  with  HST  
(Rich  et  al.\  1996;  Ajhar  et  al.,  1996;  Fusi  Pecci  et  al.,  1996;  
Holland  et  al.,  1997).    
We  have  deliberately  avoided  the  innermost  and  reddest  clusters
for  which  the  crowding  is  so  severe  as  to  compromise
photometry  in  the  visible  bands  (see  Jablonka  et  al.\  2000;  Stephens  et  al.\  2001).
  
Ten  clusters  were  observed    under  program  GO-6671  (P.I.:    Rich),    
using  the  WFPC2  on  board  HST,  and  the  filters  F555W  (V;    4  images    
of  1200,  1300,  1400,  1400  sec  on  each  cluster)  and  F814W  
(I;    4  images  of  1300,1300,1400,1400  sec  on  each  cluster).  
The  clusters  were  all  centered  on  the  PC  frame  that  provided  the  
best  spatial  resolution,  except  G91  that  fell  in  the  WF3  frame  while  
the  PC  was  pointed  on  G87.  
For  two  additional  clusters,  G302  and  G312,  that  were  observed    with  
the  same  HST+WFPC2  setup  in  the  program  GO--5906  (P.I.:  Holland),    
we  retrieved  the  data  from  the  HST  archive.  
The  total  exposure  times  are  4320  sec  (V  band)  and    4060  sec  (I  band),  
so  the  data  are comparable  with  ours.

In  Table \ref{tab:log} we give the journal of observations for these clusters,  
as well as some basic parameters. Their images in the I band are shown in 
Fig. \ref{fig:images}. 

\subsection{Data  reduction}

The  HST  frames  were  reduced  using  the  ROMAFOT  package  
(Buonanno  et  al.\  1983)  which  is  optimized  for  accurate  
photometry  in  crowded  fields,  and  has  been  repeatedly  updated
to  deal  with  HST+WFPC2  frames.  In  particular,  the  PSF  is  
modeled  by  a  Moffat  (1969)  function  in  the  central  part  of  
the  profile  plus  a  numerical  experimental  map  of  the  residuals  
in  the  wings.    The  optimal  PSF  is  determined  from  the  analysis  
of  the  brightest  uncrowded  stars  independently  in  both  
sets  of  V  and  I  co-added  frames.    

The  individual  frames  for  each  field  of  view  were  first  aligned  
and  stacked  so  as  to  identify  and  remove  blemishes  and  cosmic  ray  hits;  
this  procedure  also  made  it  possible  to  detect  and  identify  the  accurate  
positions  of  all  point  sources  including  the  faintest  ones.  
Finally,  the  photometric  reduction  procedure  was  performed  
on  the  individual  frames  and  the  instrumental  magnitudes
of  each  star  were  averaged  with  appropriate  weights.  This  procedure  
allowed  us  to  achieve  a  better  photometric  accuracy  than  
performing  photometric  measures  on  the  stacked  frames.  
No  corrections  for  non-linearity  effects  were  applied,  because  it  
was  verified  that  they  were  not  necessary.  

Individual  stars  were  measured  in  radial  annuli  whose  distances  from  
the  respective  cluster  centers  depend
on  the  intrinsic  structural  properties  of  the  GCs  and  on
the  crowding  conditions.  We  report  in  Table  \ref{tab:fsub}  for  each  observed  
cluster  the  annulus  where  photometry  was  done,  and  the  percentage  of  sampled  
light/population  over  the  total  ($L_{sam}/L_{tot}$).  
The  value  of  $L_{sam}/L_{tot}$  has  been  computed  by
integrating  the  light  in  the  annulus  directly  from  the
cluster  profile  as  obtained  in  the  study  of  the  structural
parameters  (Parmeggiani  et  al.\  2005  -- Paper  III,  in  preparation  -  
Djorgovski  et  al.\  2003).
More  internal  areas  were  too  crowded  for  individual  stellar  
photometry,  and  more  external  areas  were  dominated  by  field  
population.

The above procedure assumes that the clusters are projected spherical.
This may  be incorrect in some cases, as shown, for example, by Lupton (1989) 
and Staneva et al.\ (1996) 
using ground-based data, and confirmed by Barmby et al.\ (2002) and  
Parmeggiani et al.\ (2005) using HST data. On the basis of these HST data, we 
estimate an average ellipticity of 0.11 for all clusters except two 
(G1 and G319) for which the ellipticity is about 0.2. 
Therefore, the error we make on the $L_{sam}/L_{tot}$ ratio by assuming a 
spherical rather than elliptical light distribution is smaller than 10$\%$ 
in all cases, and is irrelevant for the present analysis.

\subsection{Calibration  and  photometric  accuracy}

The  calibration  to  standard  V  and  I  magnitudes  was  performed  according  to  
the  procedure  outlined  in  Dolphin (2000)  (updated  as  described  in  the  web
site).  
>From  this  procedure,  that  accounts  for  both  the  charge  transfer
efficiency  and  the  variations  of  the  effective  pixel  area  across  the  WFPC2,
we  have  obtained  the  final  calibrated  magnitudes  in  the  Johnson  photometric
system.

The  final    internal  photometric  errors  are  $\sim  0.02$  mag  in  (V,  I)  and  
$\sim  0.03$  mag  in  (V-I)  for  V  $<$  24.0,  and  $\sim  0.06$  mag  in  
(V,  I)  and  $\sim  0.08$  mag  in  (V-I)  for  V  $>$  25.5.  Therefore,  at  
the  level  of  the  HB  (V  $\sim$  25)  the  photometric  errors  are  typically  
$\sigma_{(V,I)}  \sim$  0.05  mag  and    $\sigma_{(V-I)}  \sim$  0.06  mag.  

The  limiting  magnitudes  of  our  photometry, defined as 5-sigma detections,  run  from  
V  $\sim$  26.2  and  
I  $\sim$  25.5  in  the  worst  crowding  case  (i.e.  G76)  to    V  $\sim$  27.2  
and  I  $\sim$  26.3  in  the  best  case  (i.e.  G11).    The  limiting  magnitude  
cutoffs  are  shown  as  dotted  lines  in  the  CMDs  presented  in  Fig.  
\ref{fig:cmd}.  These  CMDs  contain  {\em  all}  the  stars  that  were  detected  and  
measured  in  each  cluster  within  the  annuli  specified  in  Table  \ref{tab:fsub}.    

\subsection{Field  subtraction}  

For  each  globular  cluster  the  surrounding  field  was  studied  
using  the  corresponding  WFC  frames.      
These  were  reduced  with  the  photometric  package  DoPhot,  
which  runs  much  faster  than  ROMAFOT  and  equally  accurate  where  
the  crowding  is  not  too    severe.  
The  calibrations  are  compatible,  within  the  observational  errors,    
with  those  obtained  for  the  globular  clusters  using  ROMAFOT  on  
the  PC  frames.  
The  detailed  analysis  and  results  on  the  surrounding  fields  
are  presented  in  a  separate  paper  (Bellazzini  et  al.\  2003).

Our  initial  approach  used  the  outermost  annulus  in  each  PC  frame
to  define  the  field  population  for  statistical  subtraction,  but  the  field
was  so  small  as  to  lack  enough  stars  for  a  meaningful  background  sample.
Therefore  the  use  of  the  more  external  and  much  larger  
WFC  fields  seemed  preferable  as  it  provided  a  better  statistical  
base  for  field  subtraction,  once  verified  that  the  different  reduction  
packages  applied  to  the  PC  (i.e.  globular  clusters)  
and  WFC  (i.e.  fields)  frames  yield  comparable  results.    
The  statistical  field  subtraction  was  performed  using  the  
algorithm  and  the  procedure  developed  and  described  by  Bellazzini  
et  al.\  (1999b,  their  Sect.  3)  which  is  very  similar  to
that  adopted  by  Mighell  et  al.\  (1996).  As  widely
discussed  by  Bellazzini  et  al.\  (1999a),  any  procedure  aiming  at  
statistical  decontamination  suffers  of  some  degree  of  uncertainty.  
The  effects  of  decontamination  on  the  various  parts  of  the  CMD  are  
hard  to  evaluate  in  detail,  because  of  the  complex  effects  of  crowding,  
completeness  variations and background determination.
In addition, effects due to the possible existence of tidal tails can be present, 
as discussed, for example, by Grillmair et al. (1996), Holland et al. (1997), and 
Barmby et al. (2002). 
However, as noted by Meylan et al. (2001), a proper consideration of the tails would
need to reach stars a few magnitudes fainter than the turnoff to have a statistically 
significant sample of such escaping stars. This is not our case, because of the 
brighter limiting magnitudes of our photometry.  

As  an  example,  we  show  in  Fig.  \ref{fig:fieldsub}  the  CMDs  of  two  clusters,  
G76  and  G11,  that  represent  the  worst  and  best  cases  of  crowding  conditions,  
respectively.  
For  each  we  show  the  CMD  of  the  globular  cluster  before  and  after  field  
subtraction,  and  the  CMD  of  the  corresponding  subtracted  field.  As  expected,
when  the  crowding  is  low  the  field  contribution  is  nearly  irrelevant,  
and  when  the  crowding  is  high  the  subtraction  of  the  field  may  have  
significant  effects  (see  Sect.  3.2.1).          

Since  in  the  present  analysis  we  aim  at  defining  ridge  lines  and  average
properties  of  the  various  branches  of  the  CMDs,  the  statistical  decontamination  
helps  in  ``cleaning''  these  features  from  the  field  stellar  contribution,  
makes  their  detection  easier  and  highlights  the  information  there  contained.  
Therefore  we  have  applied  the  statistical  field  subtraction  to  all  our  
clusters  according  to  the  procedure  described  above.  We  list  in  Table  
\ref{tab:fsub}  the  total  number  of  measured  stars  in  each  CMD  and  
the  number  of  stars  left  after  statistical  field  subtraction,  and  we  show  
the  corresponding  field-subtracted  CMDs  in  Fig.  \ref{fig:cmdsub}.      

\subsection{Photometric  blends}  

An  inspection  of  Fig.  \ref{fig:cmdsub}  reveals  that,  in  several  
clusters,  the  HBs    appear  unusually  steep  and  heavily  populated  also  
on  the  red  side  of    the  HB,  quite  at  odds  with  the  shape  one  would  
normally  expect  by    comparison  with  Galactic  GC  of  similar  metallicity.
This  is  especially  striking  in  the  most    metal-poor  clusters  with  
normally  blue  HB  morphologies,  and  the  stars  populating
the  steep  red  HB  branch  do    not  correspond  to  any  ``classic''  
evolutionary  phase.  
The  most  plausible  explanation  is  that  these  unusual  features  are  
the  result  of  photometric  blends.  To  support  this  suggestion  
we  take  G64  as  an  example,  and  show  it  in  detail  in  Fig.  
\ref{fig:g64blend}.    In  the  top  left  panel  we  show  the  CMD  of  
the  entire  stellar  sample  we  have  measured,  after  field  decontamination.  
The  HB,  with  unquestionably  blue  morphology  in  agreement  with  the  cluster  
metallicity,  seems  to  extend  in  a  rather  steep  sequence  to  the  
bright  and  red  side,  at  V$<$25  and  0.4~$<(V-I)<$~1.0.  In  the  
other  panels  we  show  the  CMDs  of  the  stars  in  progressively  more  
external  areas  of  the  cluster,  and  we  see  that  this  feature  
becomes  less  significant  and  eventually  disappears  at  
a  radial  distance  of  about  3.2  arcsec  (i.e.  70  px).  This  
evidence  strongly  suggests  that  this    feature  is  due  to  
photometric  blends  and  not  to  real  stars. 

To test this hypothesis we have performed a simulation  based on the 
"artificial stars" method. Since the degree of crowding is highly variable 
with radial distance, we have considered three rings at 45$< r <$60 px, 
60$< r <$80 px and $r >$80 px, and determined the completeness curve for each 
of them.  
Fig.  \ref{fig:complete} reports the results of this test in G64.
As can be seen, the degree of completeness is a strong function  
of distance from the cluster center, as expected. 
From the plot one can draw two important indications: 
a) the completeness drops below 90\% at V$\sim 23.8$ (i.e. well above the HB 
level) in the innermost bin, at V$\sim 24.8$ (i.e. just above the HB level) in 
the intermediate bin, and at V$>25.8$ (i.e. fainter than the HB) in the outer ring; 
b) the fraction of recovered stars in the inner rings, and in particular in the 
intermediate one, is higher than 100$\%$. We interpret this fact as due to 
blending effects which produce more luminous stars at the expense of the fainter 
ones, thus ``drifting'' part of the fainter stellar population into an artificial 
brighter stellar population in the more internal (crowded) rings.   

As well known, if the co-added stars have different luminosities the luminosity 
and color of the blend are nearly the same as those of the brighter component; 
if the two stars have similar colors one would observe just a brightening (up to 
0.75 mag for equal components); however, if the two stars have similar 
brightness {\em and} different colors (blue and red), the resulting blend is 
brighter and with an intermediate color. 
This is exactly what has occurred with the bright red stars we are observing 
on the red HB.  To  confirm this  explanation,  we  show  in 
Fig. \ref{fig:simg64}  
how  the  simulated  combination  of  an  HB  star  fainter  than  V=25  and 
 bluer  
than  (V-I)=0.4  with  a  red  giant  star  of  similar  luminosity  
and  (V-I)$\sim$1  produces  exactly  this  type  of  feature.  
Both  blue  HB  and  red  giant  stars  are  abundant  in  metal-poor  clusters,  
and  the  extremely  high  density  conditions  in  the  innermost  areas  favor  
the  occurrence  of  photometric  blends.  

Therefore,  we  have  repeated  for  all  clusters  the  inspection  of  
the  CMDs  over  progressively  more  external  areas  to  identify  
spurious  features  due  to  blends,  if  any,  and  the  radial  distance  
at  which  these  features    become  irrelevant.  In  the  last  two  
columns  of  Table  \ref{tab:fsub}  we  list  these  values  of  radial  
distance,  and  the  number  of  stars  left  in  each  CMD  after  
subtracting  the  likely  blends.  In  the  following  analysis  we  use  CMDs  that  
are  decontaminated  from  field  and  blend  contributions,  for  a  better  
definition  of  the  CMD  intrinsic  characteristics.      
        
\section{The  CMDs:  Results  and  Analysis  of  the  Main  Branches}    

We  show  in  Fig.  \ref{fig:cmdsub}    the  CMDs  for  the  12  globular  
clusters  considered  in    the  present  study,  after  decontamination  from  
the  field  contributions.    
  
We  note  that  the  CMD  morphologies  are  generally  consistent  with  the  
metallicity  content  in  much  the  same  way  as  in  Galactic  GCs:  
(a)  the  slope  of  the  red  giant  branch  decreases  with  increasing  metallicity,  
and    (b)  the  bulk  of  the  HB  population  is  progressively  shifted  from  
the  red  in  the  metal-rich  objects  towards  the  blue
in  the  metal-poorer  ones.  

This  indicates  that  the  M31  GCs  included  in  the  present  sample
are  on  average  similar  to    those  of  the  MW.  
It  is  worth  noting  that  if  we  had  a  large  enough  sample  (say  50-60)  
of  CMDs  of  this  quality  for  the  M31  GCs,  our  knowledge  of  the  M31  GC  system  
would  be  comparable  to  what  we  knew  about  the  MW  GCs  in  the  
early  1970s.  This  level  of  knowledge  could  easily  be  achieved
with  HST.  

%
%

Relying  on  the  evidence  of  the  substantial  similarity  between  the
two  GC  systems  and  using  our  knowledge  of  Galactic  GC  as  reliable  
templates,  the  morphology  and  characteristics  of  the  two  main  
features  of  these  CMDs,  namely  the  RGB  and  the  HB,  can  be  inspected  
and  used  to  derive  information  on  a  number  of  parameters,  
that  are  described  in  detail  in  the  following  sections.  
However,  we  first  consider  the  reddening  which  has
major  impact  on  the  determination  of  parameters  such  as    
metallicity  and  distance,  as  discussed  later.  

\subsection{Individual  reddenings  from  the  literature  }

The  extinction  in  the  direction  of    M31  is  due  to  dust  that  can  reside  either  in  our  
Galaxy  or  within  M31.  
A  reliable  estimate  of  average  reddening  due  to  Galactic  material  can  
be  obtained  from  reddening  maps  (Burstein  \&  Heiles  1982;  Schlegel  et  al.\  
1998),  whereas  no  estimates  are  available  of  the  internal  reddening,  and  
this  is  one  of  the  major  sources  of  uncertainty  in  the  study  of  the  stellar  
populations  in  M31  (see  Barmby  et  al.\  2000).  

The  Galactic  reddening  in  the  direction  of  M31  was  estimated
by  many  authors:  van  den  Bergh  (1969)  found  E(B-V)=0.08,  McClure \&
Racine  (1969)  give  0.11,  Frogel  et  al.\  (1980)  0.08,  Crampton  et  al.\  (1985)  
0.10,  Jablonka  et  al.\  (1992)  0.04.  
The  maps  of  Schlegel  et  al.\  (1998)  yield  a
value  of  about  E(B-V)=0.06  in  the  direction  of  M31.  

The  reddening  of  individual  GCs  in  M31  has  been  estimated  by  several
ways,  all  of  them  quite  uncertain.  Vetesnik  (1962)  derived
color  excesses  for  257  candidate  GCs  by  assuming  an  average  true  color
for  36  GCs  located  well  outside  the  body  of  M31,  which  implied  that
these  clusters  were  only  affected  by  the  foreground  Galactic  extinction.
Later,  many    authors  adopted  the  simplified  assumption  of  a  single  intrinsic
color  for  all  GCs  in  M31.  
Frogel  et  al.\  (1980)  estimated  the  individual  reddening  for  35  GCs  using  the
reddening-free  parameter  Q$_K$  from  unpublished  spectroscopic  data  taken
by  L.  Searle.    
Crampton  et  al.\  (1985),  using  their  spectroscopic  slope  parameter
S,    derived  a  relationship  between  (B-V)$_0$  and  S  and  computed  the  intrinsic
colors  and  the  color  excesses  for  about  40  candidates  in  their  sample.
Barmby  et  al.\  (2000)  determined  the  individual  reddenings  for
314  GCs  candidates  by  assuming  that
both  the  extinction  law  and  the  GC  intrinsic  colors  are  the  same  as  in  
the  MW    and  using  correlations  between  optical  and  infrared  
colors  and  metallicity,  and  by  defining  various  ``reddening-free''  parameters.

We  list  in  Table  \ref{tab:ebmenv}  the  individual  values  of  reddening  available  
in  the  literature  and  their  sources. Some of these reddenings appear as negative values, 
just as a result of the reddening determination procedure. These values are obviously 
unphysical, and we have replaced them with the value 0.06 which represents the Galactic 
reddening in the direction of M31 according to Schlegel et al.\ (1998) maps, and hence a 
lower limit for the M31 reddening.  
The  values  we  eventually adopted for  use  in  the  following  sections (see column 12) 
are  the  unweighted  average  of  the  figures  reported  in  Table    \ref{tab:ebmenv},  
with  the  following  criteria:  
i)  the  estimates  by  Vetesnik  (1962)  and  van  den  Bergh  (1969)  were  not  used    
because  their  accuracy  is  rather  poor;  ii)  the  double  estimates  obtained  by  using  
two  sets  of  (B--V)  colors  and  the  relations  from  Crampton  et  al.\  (1985)  
(columns  8  and  9)  and  from  Barmby  et  al.\  (2000)  (columns  10  and  11)  were  
considered  only  once  each,  taking  their  respective  mean  values;  and  
iii)  no  mean  reddening  value  was  allowed  to  be  smaller  than  0.06  mag;  
when  that  happened  (i.e.  G219  and  Bo468),  the  value  0.06  was  adopted  instead.    

Some  individual  estimates  are  more  reliable  than  others,  varying  from  object  to  
object,  and  it  is  difficult  to  assess  precisely  the  error  to  associate  
to  the  final  adopted  figures.  
By  comparing  the  various  sets  of  data  from  the  different
quoted  sources  for  the  low-reddened  GCs,  a  typical  error
for  the  adopted  E(B-V)  values  should  be  about  0.04  mag,  but
it  can  well  be  larger  for  some  objects  and  smaller  for  others.  
In  Table  \ref{tab:ebmenv}  we  
list  also  the  dispersion  ($\sigma$)  values  of  our  adopted  mean  
reddening  estimates,  just  to  show  how  the  individual  estimates  can  
vary.  

As    discussed  in  Sect.  3.2.2,  such  an  uncertainty  in  the  
knowledge  of  the  individual  reddenings  may  have  quite  a  
significant  impact  on  the  determination  of  both  metallicity  and
relative  distances  based  on  the  comparison  of  the  CMDs.

\subsection{The  Red  Giant  Branch  (RGB)}

The  color  and  morphology  of  the  RGB  are  sensitive  to  metallicity,   and  its
luminosity  function  (if  sufficiently  populous  and  complete)  gives  a constraint on 
the  cluster  distance  and  stellar  evolution.  The  present  data  do  not  permit
us  to  use  either  the  RGB  tip  for  distance  determination,  or  the  RGB
bump  for  a  metallicity  constraint.

\subsubsection{The  RGB  ridge  lines}

At the bright end, cluster ridge lines suffer from the small sample size while
larger photometric errors offset the larger numbers of stars on the subgiant branch.  
For  each  cluster  the  RGB  has  been  fitted  by  a  second-  or  third-order  
polynomial  law  of  the  form  (V--I)=$f$(V).  
After  each  iteration,  stars  deviating  more  than  
2$\sigma$  in  color  from  the  best-fitting  ridge  line  were  rejected  and  the  
fitting  procedure  was  repeated  until  a  stable  solution  was  reached.
  
We can measure the  RGB  ridge  line to  $\pm$0.02  mag (color),  
except  for  the  bright end of the RGB in sparsely populated CMDs.
We  have  compared  the  ridge  lines  derived  from  the  decontaminated  (i.e.  field
and  blend  subtracted)  population  to  the  ridge  lines  derived  from  the  observed  
population  in  two  annuli  at  different  radial  distances.  
We  note  that  the  decontaminated  ridge  lines  coincide  with  the  observed    
ridge  lines  in  the  inner  annulus,  and  are  slightly  bluer  (by  $\sim$  0.02  mag)  
than  the  observed  ridge  lines  in  the  outer  annulus.  
This  effect  was  noticed  also  in  Paper  I.  
From  the  present  results  this  appears  to  be  due  to  field  contamination,  
that  has  a  stronger  effect  on  more  external  cluster  areas  and  was  not  taken  into  
account  in  Paper  I,  and  not  to  a  real  color  gradient  across  the  clusters.  
Incidentally,  the  effect  of  photometric  blends  is  not  very  important  
along  the  RGB:  whereas  the  blend  of  a  blue  and  a  red  star  would  produce  
the  feature  discussed  in  Sect.  2.5,  which  stands  out  clearly  in  the  CMD,  
the  blend  of  two  red  stars  would  produce  a  brighter  red  star  and  contribute  
to  increase  only  slightly  the  scatter  in  color,  with  no  detectable  distortion  
of  the  RGB  within  the  errors.    
  
We  list  in  Table  \ref{tab:ridge}  the  ridge  lines  we  have  derived  for  all  the  
clusters  considered  in  this  study  except  G91  for  which  a  reliable  ridge  line  
could  not  be  defined.  
A  similar  table  was  presented  for  the  8  clusters  studied  in  
Paper  I  (see  their  Table  2),  and  since  the  procedure  for  the  ridge  line
determination  is  the  same,  the  two  data  sets  are  sufficiently  homogeneous  
and  can  be  used  jointly  in  the  following  considerations.

\subsubsection{Metallicities}

In  sufficiently  old  clusters  (t~$>$~10  Gyr)  the  shape  
and  color  of  the  RGBs  depend most strongly on metallicity
and  reddening  (for  a  given  treatment  of  opacity,  convection  etc.  
this  is  reproduced  also  by  the  models).
Therefore,  in  principle,  if  either  of  these  parameters  is  known  
independently,  reliable  estimates  of  the  other  parameter      
can  be  obtained  using  the  calibrations  based  on  Galactic  GCs,  assuming
that  the  two  GC  systems  are  similar.  

In  practice,  however,  besides  the  uncertainty  in  tracing  the  RGB  
ridge  lines,  the  procedure  is  further  complicated  by obvious
issues (photometric calibration) as well as other factors (clusters dispersed 
over a 20 kpc radius would have up to 0.06 mag random distance uncertainty).  
Further issues are the calibration of color vs [Fe/H] and, finally,  
dependence on cluster composition.  Our data also are not good enough to
permit us to use the method of Sarajedini (1994) which would
simultaneously determine E(V-I) and [Fe/H].

A metallicity estimate would come either from some reddening free
parameter calibrated in terms of [Fe/H] or by assuming a value for
the reddening and comparing the dereddened cluster RGB with
a grid of calibrated ridge lines.

We  have  decided  to  apply  the  first  procedure  using  the  $S$  parameter  
defined  by  Saviane  et  al.\  (2000),  and  to compare  the  results  with  
the RGB interpolation.
We  also  anticipate  that  having  adopted  as  reference
grid  the  ridge  lines  of  Galactic  GCs  with  known  reddenings,  metallicities
and  distance  moduli,  we  apply an iterative  procedure
to  the  whole  CMD  (i.e.  including  both  the  RGB  and  HB)  which  would
yield    the  ``best-fitting  morphological''  solution  for  reddening,
metallicity  and  relative  distance  modulus,  without  making  any
{\it  a  priori}  assumption.   This latter approach is somewhat
arbitrary but has the advantage of adding constraints from the HB 
morphology.

\subsubsubsection{The  metallicity  from  the  RGB  slope}

Several  indexes
related  to  metallicity  can  be  defined,  based  on  
the  morphology  of  the  RGB  (see  Ferraro  et  al.\  1999;  
Saviane  et  al.\  2000).  Of  all  these  parameters  only  one,  the  
so-called  $S$  parameter,  is  reddening-free:  it  is  defined  as  
the  slope  of  the  line  connecting  two  points  on  the  RGB,    
the  first  one  at  the  level  of  the  HB,  and  the  second  one  2  mag  
brighter  than  the  HB.  Being  a  slope,  this  quantity  is  naturally  
independent  of  both  reddening  and  distance, depending only on 
the shape of the RGB and hence, metallicity.

Originally  defined  by  Hartwick  (1968)  in  the  (B,  B--V)
plane,  the  $S$  parameter  has  been  recently  redefined  and  
recalibrated  in  the  (V,  B--V)  plane    by  Ferraro  et  al.\  (1999)  
using  high  quality  CMDs  of  52  Galactic  GCs  collected  from  different  sources,  
and  in  the  (V,  V--I)  plane    by  Saviane  et  al.\  (2000)  using  
the  homogenous  sample  of  V,I  CMDs  for  31    Galactic
GCs  observed  by  Rosenberg  et  al  (1999).  These  parameters  are  identified  
as  $S_{2.0}$  and  $S$,  respectively.  
Ferraro  et  al.\  (1999)  provide  relations  (see  their  Table  4)  between  
their  $S_{2.0}$  parameter  and  both  [Fe/H]  (in  the  Carretta  and  Gratton  1997  
--  CG97  metallicity  scale)  and  total  [M/H]  metallicity,  and  the  rms  
error  of  their  fits  is  $\sigma$=0.18  dex.  Saviane  et  al.\  (2000)  
provide  relations  (see  their  Table  6)  between  their  $S$  parameter
and  [Fe/H]  in  both  CG97  and  Zinn  and  West  (1984  --  ZW84)  metallicity  
scales,  and  the  rms  error  of  their  fits  is  $\sigma$=0.12-0.13  dex.

We  have  measured  the  $S$  parameter  for  all  the  M31  GCs  in  our  
present  sample,  except  G91  for  which  neither  the  RGB  ridge  line  nor    
the HB  magnitude  level can  be  reliably  defined.  
In  addition,  we  have  derived  the  $S$  parameter  
for  the  6  clusters  in  Paper  I  with V  and  I  data,  and  the  $S_{2.0}$  
parameter  for  the  2  clusters  (G280  and  G351)  with  B  and  V  data.  
For  the  magnitude  level  of  the  HB,  that  enters  in  the  definition  of  
$S$,  we  have  used  the  values  described  in  Sect  3.3.2  and  
listed  in  Table  \ref{tab:vhb},  
that  have  an  average  rms  error  of  $\pm$0.1  mag.  Taking  into  account  this  
uncertainty,  the  average  error  $\Delta(V-I)=\pm$0.02  in  the  definition  
of  the  RGB  ridge  lines  (see  Sect.  3.2.1),  and  the  rms  error  0.12  dex  
of  the  calibration  fit  (Saviane  et  al.\  2000),  we  estimate  that  the  
average  rms  error  of  our  metallicity  determinations  using  the  $S$  parameter  
varies  from  $\Delta[Fe/H]\sim\pm$0.28  dex  for  the  most  metal-poor  clusters  
(e.g.  G11),  to  $\sim\pm$0.24  dex  for  intermediate  metallicity  clusters  
(e.g.  G287),  to  $\sim\pm$0.15  dex  for  the  most  metal-rich  ones  (e.g.  G319).      
Table  \ref{tab:mets}  reports  the  values  of  $S$  and  $S_{2.0}$  for  
the  entire  cluster  sample,    and  the  values  of  metallicities  we  have  derived  
using  the    Saviane  et  al.\  (2000)  and  Ferraro  et  al.\  (1999)  relations,  
as  appropriate.  

\subsubsubsection{The  metallicity  from  comparison  of  the  RGB  with  
template  ridge  lines}

As  was  done  in  Paper  I  for  8  GCs  and  in  Bellazzini  et  al.\  (2003)  
for  16  fields,  metallicities  can  be  estimated  by  comparing  directly  the  
target  RGBs  with  a  reference  grid  of  Galactic  GC  (GGC)  fiducials  
of  known  metallicity,  after  correcting  for  the  respective  reddenings  and  
distances.  The  accuracy  of  this  procedure  depends  mostly  on  how  finely  
the  metallicity  range  is  sampled,  as  well  as  on  the  accurate  knowledge  of  
the  reference  grid  relevant  parameters  (i.e.  reddening  and  distance).    
  
In  order  to  check  the  results  derived  above  with  the  $S$-parameter,  
we  have  applied  to  our  M31  GCs  the  same  interpolating  procedure  used  
by  Bellazzini  et  al.\  (2003).  
The  reference  grid  of  GGCs  and  their  parameters  are  listed  in  
Table  \ref{tab:grid}.  
The  V  and  I  photometric  data  used  to  derive  the  HB  and  RGB  ridge  lines  of  
the  reference  clusters  have  been  taken  from  Rosenberg et al.\ (2000a,b) and 
Guarnieri et al.\ (1998) 
except  for  the  clusters  G280  and  G351,  that  have  HST-FOC  B  
and  V  data.  For  them,  the  template  HB  and  RGB  ridge  lines  were  derived  from  
the  B,V  database  of  GGCs  collected  by  Piotto  et  al.\  (2002).  
  
We  show  in  Fig.  \ref{fig:rgb1}  and  \ref{fig:rgb2}  the  CMDs  of  our  M31  GCs,  
plotted  individually  along  with  the  grid  of  HB  and  RGB  template  ridge  lines,  
for  comparison.  For  the  M31  clusters  we  used  the  CMDs  that  had  been  previously  
cleaned  of  field  and  blend  contamination,  and  corrected  them  for  reddening  and  
absorption  using  the  relations  E(V--I)=1.375E(B--V),  A$_V$=3.1E(B--V)  and  
A$_I$=1.94E(B--V)  (Schlegel et al.\ 1998)   
and  the  adopted  reddening  values  
listed  in  Table  \ref{tab:ebmenv}  (column  12)  as  initial  input  values.  
Then  each  target  cluster  CMD  is shifted  in  magnitude  until  reaching a  satisfactory  
match  with  both  HB  and  RGB  ridge  lines  of  a  template  CMD,  or  of  an  
``interpolated''  solution  between  two  bracketing  templates.
The  metallicity  of  the  template,  or  the  
intermediate  value  between  the  bracketing  templates  if  interpolation is
needed,  is the adopted metallicity.
The  accuracy  of  these  estimates  is  typically  half  the  
interval  of  the  bracketing  templates, about  $\pm$0.15-0.20  dex  at  
the  most  metal-poor  end  of  the  metallicity  range,  and  $\le$0.1  dex  at  the  
metal-rich  end.  
These  values  are  listed  in  Table  \ref{tab:mets}  col.  6  (labelled  $Ridge$).  

This  procedure  also  yields  an  estimate  of  distance  via  the  magnitude  shift  
that  needs  to  be  applied  in  order  to  match  the  target  CMDs  with  the  templates.    
In  some  cases  an  additional  shift  in  magnitude  with  respect  to  the  
average  distance  assumed  for  M31 is necessary  to  achieve  an  acceptable  
match.   If  no  color  shift  is  involved  as  well,  this  can  only  be  interpreted  
as  a  distance  effect,  indicating  that  the  cluster  distance  is  larger  or  
smaller  than  the  distance we have adopted  for  M31, i.e.  (m--M)$_0$=24.47$\pm$0.03    
(weighted mean of the most recent determinations from  Holland 1998; Stanek \& Garnavich 
1998; Freedman et al.\  2001; Durrell et al.\  2001; Joshi et al.\  2003; 
Brown et al.\  2004a; McConnachie et al.\  2004).   
We  remind  the  reader  that  a  dispersion  
of  $\pm$0.06  mag  in  the  distance  moduli  can  well  be  intrinsic  if  our  
clusters  are  located  on  a  spheroidal  distribution  with  $r\sim$20  kpc.  
We  have  listed  these  distance  moduli  in  Table  \ref{tab:ebvmod}  along  
with  the  corresponding  values  of  (adopted)  reddening  and  (derived)  metallicity  
(columns  3-5),  for  the  sake  of  convenience  when  we  discuss  the  issue  of  
distance  estimates  (Sect.  3.3.2).

To get the largest possible sample,  we apply this procedure to  all  
available  M31 GCs including those of Paper I.
As  shown in  Fig.  \ref{fig:rgb1}  and
\ref{fig:rgb2}  we get a satisfactory match in most cases, but it is also evident that some CMDS (e.g. G322) require an additional color shift
before they match the Galactic fiducials. This kind of problem leads us
to consider one final method below.

\subsubsubsection{An  alternative  experiment:  a  global  
``best  morphological  match''  with  the  reference  grid}

Before comparing our photometric metallicities with other approaches,
we compare M31 to MW clusters using yet a different method,
required by the few CMDs which fail the aforementioned grid because
they require a color shift.  The aim is search for the best match of both
the RGB and HB while leaving distance and reddening as free parameters.

The  values  of  metallicity  thus  derived  are  listed  in  Table  \ref{tab:mets},  
where  we  collect  all  available  estimates  of  metallicity  for  our  clusters.  
The  set  of  values  for  reddening,  metallicity  and  distance  that  yield  the  
best  match    are  reported  in  Table  \ref{tab:ebvmod},  columns  6-8.

\subsubsection{Final  considerations  on  reddening    and  metallicity}

Considering  the  results  obtained  in  the  previous  sections,  it  is  now  possible
to  discuss  these  parameters  in  some  more  detail.

\noindent{\bf  Reddening}\\
A  comparison  of  the  figures  reported  for  each  cluster  in  Table  \ref{tab:ebvmod}
(columns  3  and  6)  shows  the  difference  between  the  adopted  value
derived  in  Sect.  3.1  as  the  mean  of  the  available  estimates  in  the  literature,  
and  the  value  obtained  by  the  ``best  3-parameter  match''  of  the  whole  CMD.  
The  two  values  agree generally within  the  estimated  error  of  
$\pm$0.04  mag except  for  a  few  clusters.
The  clusters  for  which  $\Delta  E(B-V)>$~0.04  mag  are  G33,  G108,  G280,  
G302,  G319  and  G322.  

Since  we  do  not  have  any  external  strong  constraint  on  metallicity
and  distance  modulus  which  might  clarify  the  choice,  we  keep    
as  the  most  probable  values  of  reddening  those  adopted  in  Sect.  3.1  and  
listed  in  Table  \ref{tab:ebmenv},  column  12  (reported  also  in  Table  
\ref{tab:ebvmod},  column  3),  recalling  however  the  caution  implied  by  the  
worse  global  fit.        

\noindent{\bf  Metallicity}  \\  
In  Table    \ref{tab:mets}  we  list  all  the  available  independent  estimates  of  
metallicity,  for  ease  of  comparison.
In  addition  to  those  obtained  from  the  $S$-parameter  (col.s  4  and  5)  and  
from  the  two  CMD-fits  (col.s  6  and  7),  there  are  three  further  determinations,  
two  of  which  based  on  spectra  and  one  on  optical  and  IR  photometry.  

The  spectroscopic  estimates are from  calibrations  applied  to  spectral  
line  indices  in  the  cluster  integrated  spectra.  In  particular  we  report  the  data
from  Huchra  et  al.\  (1991),  collected  by    Barmby  et  al.\  (2000)  and  Perrett  et  al.\  
(2002).  

Photometric  estimates use integrated  (V--K)  colors  calibrated  
in  terms  of  [Fe/H]  in  the  ZW  metallicity scale  by  Bonoli  et  al.\  (1987).    
The  photometric  estimates  require  the  knowledge  of  the  individual  reddenings,  
which  were  adopted  by  Bonoli  et  al.\  (1987)  as  E(B--V)=0.10  for  all  the  considered  
clusters  except  G33  and  G64,  for  which  0.22  and  0.17 were  used,   
respectively.  


These    different  techniques  give  a  range  in  accuracy  and  
reliability.  
For  the  old  clusters  the  metallicities  derived  from  the  spectra  are  probably  the  
most  reliable  in  spite  of  the  uncertainties and ambiguities related  
to  the  definition,  meaning  and  calibration  of  the  spectroscopic  
indexes  used  for  these  determinations  (Burstein  et  al.\  2004  and  references  therein).  

On  the  other  hand,  the  results  from  the  ridge  line  fitting  method  are  mostly  
qualitative  and  generally  give  only  a  rough  consistency  check.

The  $S$-parameter  and  the  integrated  IR  photometry  do  not  measure  metallicity  
directly,  but  rely  on  some  type  of  calibration  which  may  introduce  additional  
uncertainty.  However,  they  are  quantitative  and  relatively  accurate  methods  
and  should  yield  quite  reliable  results.  Incidentally,  we  note  that  the  
$S$-method  depends  only  on  the  RGB  morphology,  as  well  as  the  integrated  IR  
photometry  that  is  obviously  mostly  sensitive  to  the  RGB  stellar  population,  
whereas  spectroscopic  metallicities  are  based  on  integrated  visual  spectra  
that  may  be  affected  by  other  types  of  stellar  populations  than  the  RGB,  if  
sufficiently  abundant  or  luminous.    

We are pleased that for most clusters,
the metallicity estimates we obtain from the RGBs, both via the $S$ paramter and from
direct comparison with RGB templates agree well with the other estimates.

From the above methods we derive mean values of metallicity for our clusters using
a straight unweighted average of the values obtained from the $S$ parameter,
RGB ridge line fit, spectroscopic estimates, and IR photometric values.
We  give this as a summary list in  Table  \ref{tab:mets},  
and    we  report  them  also  in  Table    \ref{tab:vhb}  for  convenience,    
since  we  use  them  in  the  following  sections.  In  Table  \ref{tab:mets}  we  
report  also  the  dispersion  ($\sigma$)  of  these  estimates which sometimes
is very small.
However,  these  are  not  the  errors  to  be  associated  to  the  final  
adopted  values.    We  estimate  that  a  realistic  error  on  metallicity  is  about  
$\pm$0.2  dex.


\subsection{The Horizontal  Branch  (HB)}

As  we  have  seen  e.g.  in  Fig.  \ref{fig:cmd},  the  HB  morphologies  are  
generally  similar  to  those  of  GGCs,  the  first  parameter  
being  metallicity.  However,  as  clearly  shown  in  Fig.  \ref{fig:cutoff},  
we  note  that  the  magnitude  limits  of  our  photometry  would  not  allow  us  
to  detect  extended  blue  tails  that  might  reach  as  faint  as  $\sim$ 3 mag  
below  the  HB  magnitude  level.  
For  example,  a  cluster  such  as  NGC  6752,  where  the  extended  blue  tail  
reaches  as  faint  as  the  TO  and  contains  a  significant  fraction  of  the  total  
HB  stellar  population, would be measured as having a blue HB even though
half such stars are undetected.

In  order  to  test  how  the  photometric  cutoffs  affect  the  
appearence  of  the  HBs,    we  have  taken  the  
well  known  and  accurate  CMDs  of  6  of  the  best  studied  GGCs,  
namely  M3,  M13,  M92,  47  Tuc,  NGC6752  and  NGC2808,  and  we  have  
shifted  the  limiting  magnitude  cutoffs  to  the  assumed  distance  
to  M31.    The  individual  GGC  reddening  values  have  been  taken    
from  Harris  (1996,  update  2003).  
We  have  selected  these  clusters  because  they  
represent  good  cases  of  low,  intermediate  and  high  metallicity,  
of  extended  blue  HB  tail,  and  of  a  2nd-parameter  HB  morphology.    
In  Fig. \ref{fig:cutoff}  we  show  the  results  of  this  simulation.    
We  note  that  a  significant  part  of  the  HB  population  is  lost  
when  the  HB  is  very  blue  and  extends  down  to  magnitudes  as  faint  
as  the  MS  turnoff.  

Therefore any conclusions based  on  the  HB  
morphology  must necessarily  be qualitative.
For  example,  we  cannot  hope  to  estimate  
the  helium  abundance  $Y$  of  our  clusters  using  the  $R$-method  
(Buzzoni  et  al.\  1983),    because  this  method  relies  on  stellar  
counts  in  the  RGB  and  HB  evolutionary  phases  and  completeness  
is  an  essential  requirement.      
  
However,  more  qualitative  considerations  are  possible,  for  example  
we  note  that  the  HB  morphologies  of  G119  and  G105    
are  quite  different  even  though  they  have  very  similar  metallicity.
This  suggests  the  presence  of  2nd-parameter  clusters  among  
our  M31  clusters,  as  occurs  in  the  MW  GC  system  and  in  the  Fornax
dwarf  spheroidal  galaxy  (Smith  et  al.\  1996; Buonanno et al.\ 1998).

Quantitative  considerations  can  also  be  possible,  provided  they  are  
not  affected  by  the  magnitude  cutoffs.  For  example,  the  magnitude  level  
of  the  HB,  V(HB),  at  the  expected  position  of  the  instability  strip    
(i.e.  at  approximately  0.3~$\lesssim  (V-I)_0  \lesssim$~0.7),  can  be  
estimated  with  a  good  level  of  confidence  and  accuracy.  
These  are  the  features  that  we  analyse  and  discuss  in  the  following  
sections.

\subsubsection{The  HB  morphology:  2nd-parameter  (2nd-P)  effect}

The  morphology  of  the  HB  depends  primarily  on  [Fe/H],  
metal-poor  (-rich)  clusters  having  predominantly    blue  
(red)  HBs.  However  in  the  Galaxy  there  are  several  cases  
where  this  general  rule  is  not  followed  and  the  presence  
of  a  second  (or  more)  parameter(s)  must  be  invoked
(see  for  references    Fusi  Pecci \& Bellazzini  1997).  

The  2nd-P  candidate  that  has  been  most  often  suggested  is  age:  
for  a  given  metallicity,  age  affects  the  HB  morphology  in  the  sense  
that  older  clusters  have  bluer  HBs.  However,  at  least  in  the  case  
of  NGC  2808,  where  the  color  distribution  of  the  HB  stars  is  bimodal  
and  appears  to  be  the  sum  of  NGC362  and  NGC288,  the  2nd-P  seems  to  be  at  
work  within  the  cluster  itself.  Therefore  gross  age  differences  alone  are  not
responsible  for  this  unusual  HB  (unless,  of  course,  one  
assumes  that  the  cluster  contains  at  least  two  different  generations  of  
stellar  populations,    see  for  references  D'Antona  and  Caloi  2004;  also  note
the  role  of  helium  abundance  as  in  $\omega$  Cen;  Piotto  et  al.\  2004).    

A common approach to calculating a morphology derived HB index uses
(B-R)/(B+V+R),  where  V  indicates  the  number  of  variable  HB  stars  
(i.e.  RR  Lyraes)  within  the  instability  strip,  and  B  and  R  the  number  of  
HB  stars  bluer  and  redder  than  the  instability  strip,  respectively  
(Lee  et  al.\  1994).  Lacking a variable star census, we use
the  Mironov (1972) index,   B/(B+R),  where  the  
boundary  between  the  blue  (B)  and  red  (R)  part  of  the  HB  is  set  at  
V-I  =  0.50.  
When  the  stellar  distribution  along  the  HB  is  known  with  relatively  poor  
accuracy  and  there  is  no  knowledge  of  the  variable  stellar  component,  as  is  
the  case  for  our  M31  GCs,  the  Mironov  index  is  quite  adequate  to  describe
the  HB  morphology.  We  list  in  Table  \ref{tab:vhb}  the  values  of  the  Mironov  
index  we  have  estimated  for  our  M31  GCs.    

We  show  in  Fig.  \ref{fig:hb}  the  behavior  of  the  Mironov  index  versus  [Fe/H]  
(both  taken  from  Table  \ref{tab:vhb})  for  our  GCs  in  M31  (shown  as  filled  
circles  and  identified  with  their  names).  For  comparison,  we  show  the  
GGCs  for  which  the  Mironov  index  is  available  (small  open  circles),  and  their  
general  behavior  as  a  shaded  area  whose  mean  line  is  drawn  in  analogy  with
the  approach  described  by  Lee  et  al.\  (1994)  for  (B-R)/(B+V+R).    
As  is    well  known,  in  the  MW  the  2nd-P  phenomenon  affects  only  
clusters  in  the  metallicity  range  approximately  between  [Fe/H]=--1.1  and  
--1.6.

The  two  cluster  systems  behave  in  a  roughly  similar  way,  except
that  the  spread  in  HB  type  for  the  metal  poor  ($\rm  [Fe/H]\sim  -1.7$) 
clusters  is  striking,  suggesting  an  offset  from  the  Milky  Way  trend  line.
An  underestimate  of  the  B  counts  due  to  the  loss  of  extended  HB
stars  may  explain  part  of  this  second  parameter  effect.
However,  G87,  G287,  and  to  some  extent  G11  and
G219  appear  to  genuinely  display  the  second  parameter  effect.
As  good  candidates  for  the  2nd-P  effect,  we  also  note  the  pair  G105  and  G119,  
that  have  very  similar  metallicity:  G105  falls  nicely  on  the  distribution  
defined  by  the  GGCs,  whereas  G119  presents  a  distinctly  redder  HB  morphology.  

The  Fornax dwarf  galaxy  shows  the  second  parameter  problem  but  at  
even  lower  metallicity    than  the  M31  halo  (Smith et al.\  1996).
The  evidence  for  an  age  dispersion  in  the  M31  halo  field  population  
(Brown  et  al.\  2003) opens  the  question  of  whether the metal poor 
populations might have an age dispersion  (note  however  that  the  strongest  
evidence for an intermediate age component is in the metal rich population).
Buonnano  et  al.\  (1998)  report  deep  HST  photometry reaching  the  
turnoff  point  and place an  age  spread  of  $<$1  Gyr  on  the  Fornax
globular clusters;  they  further  claim  that the
Fornax  globular  clusters  have  the  same  
age  as  the  oldest  Milky  Way  clusters  (e.g.  M92).      
It  would  be  valuable  to  undertake deeper  observations  of  M31  globular  
clusters,  at  higher  resolution,  to  explore  the  question  of  the  second  
parameter  problem  and  the  relationship to  the  Milky  Way  clusters.      
Evidence  for  a  systematic  difference  in  the
second  parameter  effect  between  the  Milky  Way,  the  Fornax  dwarf,
and  M31  could  point  toward  differences  in  age  or  chemistry.    It  has
been  argued  that  the  Local  Group  experienced  a  common  era  of  star
formation  as  evidenced  by  the  nearly  identical  age  of  the  old
halo  globular  clusters  (see  for  example  Harris  et  al.\  1997;  Rosenberg  et  al.\  2002).

We  may  conclude  that  M31  exhibits  the  second  parameter  effect  at  lower  metallicity
than  does  the  Milky  Way,  but  not  in  extreme  a  sense  as  is  evidenced  in  the  Fornax
dwarf.

\subsubsection{The  HB  Luminosity-Metallicity  Relation}

The  Population  II  distance  scale  is  based  on  the  absolute  magnitude  of  local
calibrators,  i.e.  the  RR  Lyrae  stars,  M$_V$(RR),  which  is  known  to  be  
dependent  on  metallicity.  This  dependence  has  often  been  represented  by  a  
linear  relation    of  the  form  $M_V  =  \alpha  {\rm[Fe/H]}  +  \beta$,  which  seems  
to  represent  the  observed  behavior  of  these  stars  reasonably  well  
(within  the  uncertainties),  although  some  theoretical  pulsation  and  evolution  
models  suggest  that  a  non-linear  (quadratic?)  relation  might  be  more  
appropriate.  Given  the  uncertainties  of  our  estimates  in  the  M31  GCs,  
the  linear  relation  is  quite  adequate  to  our  considerations.  We  refer  the  
reader  to  Cacciari  \&  Clementini  (2003)  for  a  recent  and  detailed  review  on  
this  subject.  

After  a  lengthly  period  of  debate,  $\alpha$  appears    
to  be  converging  towards  a  value  of  
$\sim$0.20-0.23:  for  example  Gratton  
et  al.\  (2004)  find  0.214$\pm$0.047  mag  from  the  analysis  of  98  RR  Lyrae  stars  
in  the  bar  of  the  LMC,  in  general  agreement  with  several  recent  independent  
estimates.  

Also  disagreement  on  the  zero-point  of  this  relation  are  
finding  a  solution:  once  the  most  extreme  (and  less  reliable)  determinations  
are  taken  into  account  with  the  proper  weight,  the  most  recent  and  accurate  
results  seem  to  converge  on  two  sets  of  values  that  differ  only  by  
$\lesssim$0.05  mag.  
The  ``faint''  solution  converges  toward  $M_V$(RR)=0.59  mag  at
[Fe/H]=--1.5,  and  is  mainly  supported  by  the  result  of  the  HST  trigonometric  
parallax  on  RR  Lyr  (Benedict et al.\ 2002) which we consider doubtful
because  of  its    rms  error
of  $\pm$0.1  mag.  
On  the  other  hand,  the  ``bright'      solution  converges  toward  
$M_V$(RR)=0.55  mag  at  [Fe/H]=--1.5,  and  is  mainly  supported  by  the  results  
of  various  studies  on  the  LMC  distance  scale  which  are  discussed  and  
summarised  as  (m--M)$_0$(LMC)=18.515$\pm$0.085  by  Clementini  et  al.\  (2003).  

Theoretical  evolution  and  pulsation  models  agree  with  either  estimate,  within
the  respective  errors.  

We  can  hardly  say  there  is  any  discrepancy  left  at  all,  therefore  we  assume  
that  a  relation, 

\noindent $M_V  =  0.22{\rm[Fe/H]}  +  0.88$  \hfill (1)

\noindent with  an  rms  error  $\pm$0.05  mag,
represents  the  behavior  of  the  RR  Lyrae  stars  quite  reliably,  and  
accurately  enough  for  use  in  our  M31  GCs.    This  relation  corresponds  to  
a  distance  to  the  LMC  of  (m--M)$_0$(LMC)=18.51  using  Clementini  et  al.\  (2003)
data.   We  use  it  later  to estimate $M_V$ and hence the distance 
(see case [C] in Table \ref{tab:vhb}). 
    
The  CMDs  of  our  M31  globular  clusters,  reaching  about  one  magnitude  below  
the  HB,  offer  an  interesting  means  to  test  the  characteristics  
of  the  RR  Lyrae  stars  in  another  galaxy  (via  the  slope  of  the    
$M_V$(HB)-[Fe/H]  relation)  and  derive  the  distance  to  M31,  or,  alternatively,  
to  yield  quantitative  estimates    on  the  spatial  distribution  of  our   
clusters  within  the  M31  spheroid.  


\subsubsubsection{The  observed  HB  magnitude  level  V(HB)}

We  have  estimated  the  apparent  mean  magnitude  of  the  HB,  V(HB),  by  a  running  
box  averaging  method  and  adopted  the  V(HB)  value  at  (V--I)$_0$=0.5,  
corresponding  to  the  mid  point  of  the  instability  strip  that  covers  the  range  
of  (V--I)$_0$  colors  approximately  between  0.3  and  0.7.  

For  the  metal-rich  clusters  ([Fe/H]$>$--1.0)  where  only  the  red  HB  clump  is  
detected,  we  have  used  the  $\langle  V\rangle$  of  the  red  HB  clump  corrected  by    
+0.08  mag  (see  Paper  I  and  references  therein).    
We  list  the  values  of  the  observed  V(HB)  magnitudes  in  Table  \ref{tab:vhb}.  

Typical  photometric  errors  on  the  individual  HB  stars  are  about  
$\pm$0.05  mag  (see  Sect.  2.3),  that  become  $\pm$0.02-0.01  mag  when  the  
average  value  V(HB)  of  10-25  stars  is  taken.  
Considering  an  additional  error  of  about  $\pm$0.04  mag  on  the  
reddening  values,  the  typical  rms  error  we  associate  to  V$_0$(HB)  is  
$\pm$0.13  mag.

\subsubsubsection{M$_V$(HB)  and  the  distance  to  M31}

We  recall  that  in  Sect.  3.2.2,  as  a  result  of  the  application  of  
the  RGB  ridge  line  and  global  fitting  methods  (with  fixed  or  free  value  
for  the  reddening),  we  have  derived  for  each  cluster  two  estimates  of  
distance  (see  Table  \ref{tab:ebvmod}  columns  5  and  8,  reported  
also  in    Table  \ref{tab:vhb}  as  cases  [A]  and  [B],  for  the  sake  of  
convenience  in  the  following  discussion).  
These  two  sets  are  substantially  compatible  within  the  errors,  
with  the  exception  of  G33,  G76,  G302  and  G322  where  the  two  estimates  differ  
by  $\ge$0.15  mag,  mostly  because  of  significantly  different  values  
of  reddening.  Both  methods  yield  an  average  distance  modulus  
(m--M)$_0$(M31)=24.48$\pm$0.12  mag.    

It  is  important  to  note  that  the  values  so  obtained  for  M$_V$(HB)  cannot  
be  used  to  derive  an  independent    M$_{V}$  versus  [Fe/H]  relationship,  as  they  
actually  reflect  the  relationship  adopted  by  Harris  (1996)  to  determine  the  
distances  of  the  GGCs  used  in  the  reference  grid,  namely    
$M_{V}  =  (0.15\pm0.05){\rm[Fe/H]}  +  (0.76\pm0.07)$.  They  can  be  used,  however,  
to  get  some  hint  on  the  cluster  locations  and  relative  distances,  in  
particular  if  the  distance  turns  out  systematically  and  significantly  larger  
or  smaller  than  the  assumed    value  for  M31.  

However,  there  are  other  ways  of  deriving  useful  information  on  the  
M$_V$(HB)  versus  [Fe/H]  relation,  once  the  values  for  reddening,  metallicity  
and  HB  apparent  luminosity  are  known.    
In  particular,  we  can    derive  an  independent  slope  for  the  M$_V$  versus  
[Fe/H]  relationship,  
the  zero-point  depending  on  an  assumed  value  of  distance  to  M31.

By  adopting  the  same  distance  to  all  the  M31  clusters,  i.e.    
(m--M)$_0$(M31)=24.47$\pm$0.03  (see Sect. 3.2.2)  and  the  values  of    
reddening,  metallicity  and  V(HB)  reported  in  Table  \ref{tab:vhb}  columns  3,  4  
and  5,  one  can  derive  the  corresponding  values  for  M$_{V}$  (reported  under  
case  [D]  in  Table  \ref{tab:vhb})  and  the  slope  of  the  M$_{V}$  versus  [Fe/H]  
relationship  independent  of  any  other  assumption.  
Since  we  assume  that  all  clusters  are  at  the  same  distance,  the  possible  
distance  dispersion  shows  up  as  a  larger  dispersion  in  the  M$_{V}$  values.    

In  Fig.  \ref{fig:mvfe}  we  show  the  present  sample of GCs  (filled  circles),  
along  with  the  8  GCs  that  were  studied  in  Paper  I  and  reanalysed  here  
(open  circles).  
The  error-weighted  least  squares  linear  fit  to  these  data,  assuming  
errors  on  both  [Fe/H]  ($\pm$0.2  dex)  and  M$_V$(HB)  ($\pm$0.13  mag),  is  
shown  in  the  figure  and  yields: 

\noindent $M_V = (0.20\pm0.09){\rm[Fe/H]} + (0.81\pm0.13)$ \hfill (2)     

\noindent As noted above, the large dispersion of the M$_{V}$ values is due to the  
combination  of  two  effects:  i)  the  photometric  errors  and  the  uncertainties  
in  the  method  applied  to  determine  the  average  V(HB),  and  ii)  the  intrinsic  
luminosity  dispersion  due  to  the  relative  distances  of  the  clusters  within  M31.
    
Concerning  the  slope  of  eq. (2),  we  note  that  the  value  0.20  
is  in  excellent  agreement  with  the  best  estimates  presently  
available  (see  eq.[1]),  based    on  Galactic  and  LMC  field  RR  Lyraes.  
We  point  out  that  the  relatively  low  value  of  $\alpha  \sim$  0.13,  that  was  
derived  in  Paper  I  from  the  analysis  of  the  first  8  M31  clusters,  was  
obviously  affected  by  large  inaccuracies  due  to  the  very  small  sample  size.  
That  result  is  now  revised  with  the  addition  of  our  new  data  and  the  
re-evaluation  of  the  clusters  studied  by  Holland  et  al.\    (1997),  that  have  
more  than  doubled  the  previous  sample.  
As  for  the  zero-point  of  eq.  (2),  it  obviously  depends  on  the  assumption  we  
have  made  on  the  distance  to  M31  that  leads  to  M$_{V}$(HB)=0.51  at  
[Fe/H]=--1.5,  corresponding  to  (m--M)$_0$(LMC)$\sim$18.55.   

This finding leads us to conclude that the HB  stars  in  M31  
do  indeed  share  the  same  physical  behavior  as  in  the  MW  and  in  the  LMC,  
whether  they  belong  to  the  field  or  to  the  GC  stellar  population.    
In  particular,  their  luminosity  varies  as  a  function  of  metallicity  in  much  
the  same  way.  

If  we  now    assume  that  there  is  one  ``universal''  M$_{V}$  versus  [Fe/H]
relation,  for  example  eq. (1) that we  have  derived  before based on Galactic and LMC  
RR  Lyraes,  and  calibrated  on  (m--M)$_0$(LMC)=18.51,  we  can  derive  the  
individual  values  of  $M_V$(HB)  irrespective  of  their  distances.  
So  doing,  the  dispersion  due  to  the  cluster  location  within  the  M31  
spheroid  now  shows  up  through  the  derived  distance  moduli  and  not  $M_V$(HB)  
(see  case  [C]  in  Table  \ref{tab:vhb}).  The  straight  
average  value  of  the  derived  distance  moduli  in  case  (C)  is  24.44$\pm$0.19  
considering  all  19  clusters,  and  24.47$\pm$0.10  considering  only  the  14  
clusters  whose  modulus  deviates  by  less  than  0.2  mag  (i.e.  1$\sigma$)  
from  the  average.  
If we compare these distance moduli with those determined by Holland (1998) by fitting 
theoretical isochrones to the observed RGBs of 14 GCs, we note that the mean values 
are identical, i.e. 24.47. However, the {\em individual} values for the 11 clusters in 
common can differ randomly by up to 3$\sigma$ in a few cases. 
Considering that the two sets of results are based on different assumptions (e.g.\ on 
reddening and metallicity) and different methods (i.e.\ fitting the observed RGBs to 
theoretical isochrones instead of using template ridge lines), these differences are 
quite acceptable and point out the {\em true} uncertainties of these determinations. 

\noindent  From  this  exercise  we conclude:

\noindent  i)  The  mean  distance  to  M31  agrees with  the assumed  value 
(m--M)$_0$(M31)=24.47  well within the error determinations, and consistently with 
(m--M)$_0$(LMC)=18.51-18.55.  

\noindent  ii)  Several of  our  target  clusters  appear  to  lie  off  the  assumed  
spheroidal  distribution  of  radius  $\sim$20  kpc  (0.06  mag  in  distance  modulus),  
i.e.  at  larger  distances  from  the  galactic  center  along  the  line  of  
sight.  Our measures  indicate  that  G64,  G76  and  G322  are  located  on  the  
near  side  of  the  spheroid,  whereas  G105,  G351  and  possibly  G108  and  G219  are  
located  on  the  far  side.

\section{Summary  and  Conclusions}  

We  have  presented  the  results  of  our  HST+WFPC2  observations  in  the  filters
F555W  (V)  and  F814W  (I)  for  10  globular  clusters  in  M31,  and  the  
reanalysis  of  2  more  clusters  using  HST  archive  data  of  comparable  
characteristics.  
    
We  obtain  high  quality  CMDs  down  to  approximately  1  magnitude  
below  the  HB.   The principle sequences (HB and RGB) look similar to
those seen in old Milky Way globular clusters.

We include  also in our sample the  CMDs  for  8  clusters  
previously analyzed in  Paper  I,  for  a  more  general  and
comprehensive  discussion  of  the  M31  GC  system  characteristics;
we conclude as follows:

1. We  derive  metallicities  from  the  RGB  ridgelines;  these  are  in  good  agreement
with  those  derived  from  integrated  ground-based  spectroscopic  and/or  photometric  
estimates.
  
2.  The  HB  morphologies  show  the  same  behavior  with  [Fe/H]  as  in  the  Milky  
Way,  including  the  possible  presence  of  some  2nd-parameter  clusters.    An  
apparently peculiar  HB  morphology  (bright  red  stars  and  a  slanting  HB)  is  shown  
to be  due  to  blends  of  blue  HB  and  RGB  stars.    We  also  correct  for  field
contamination,  when  this  is  an  issue.    We  observe  the  2nd  parameter  effect  at  
[Fe/H]=--1.6,  more  metal poor  than  is  seen  in  the  Galaxy,  causing  the  trend  of  
HB-type  with  [Fe/H] to  be  offset. A  possible  explanation  for  some  of  the  
metal-poor  red-HB  clusters  that  deviate  most  from  the  analogous  distribution  of  
GGCs  is  that  they  are  a  few  Gyr  younger,  as  estimated  for  G312  
(Brown  et  al.\  2004b).  
  
3.  The  mean  magnitude  of  the  HB  at  the  approximate  location  of  the  instability  
strip  has  been  estimated,  and  a  M$_V$(HB)  versus  [Fe/H]  relationship  has  
been  derived.   The  range  of  metallicities  spanned  by  the  clusters  make  it  possible
to  derive  the  slope  of  the  M$_V$(HB)  versus  [Fe/H] relationship. We  find
this  slope  to  be  $\alpha=0.20\pm0.09$, in excellent  agreement  with  independent  
estimates  based  on  Galactic  and  LMC  RR  Lyrae  stars.  
This  distance  scale, based on  (m--M)$_0$(M31)=24.47, is consistent  
with  (m--M)$_0$(LMC)=18.55.

4.  Relative  distances  could  be  estimated,  and there is evidence that  a  few  clusters   
lie   on  the  foreground  or  background  of  the  M31  main  body.

Our  results  add  further  support  to  previous  conclusions  that  the  GC  systems  
in  the  Galaxy  and  in  M31  are  basically  very  similar.  However,  a  sample  of  
19  clusters  represents  less  than  5\%  of  the  total  cluster  population  in  M31,  
and  no  firm  conclusions  can  be  drawn  from  such  a  sample,
especially  concerning  relatively  rare  objects  such  as  the  2nd-P  clusters,  
or  the  possible  presence  of  a  younger  population.    Subtle differences between
the Milky Way and M31 might follow from differing formation times or chemical
evolution, but with such a small sample of M31 clusters, it is likely we are missing
many interesting and crucial examples.  A larger sample would place our observational
description of the M31 cluster system at the same level as that of the Milky Way globular
clusters just prior to the advent of CCD photometry.  
A larger sample would also give insight into the differences between the Milky Way and 
M31 that follow from possibly different ages and chemical evolution, essential qualities 
for a better understanding of the formation and evolution of galaxies like our own.

\acknowledgements  

We wish to thank G. Parmeggiani and E. Diolaiti for help in preparing 
Figure 1, and S. Galleti for help with the artificial star simulation.   
RMR and SGD acknowledge the NASA grant n. GO-6671. 
CCE, CC, LF and FFP acknowledge the support of the Ministero dell'Istruzione, 
dell'Universit\`a e della Ricerca through the grants ASI J/R/35/00 and 
MIUR MM02241491-004.


\begin{table*}
\begin{center}
\caption{The sample of M31 globular clusters observed with the HST and analysed  
in the present study. The data for the first 10 clusters are from the program  
GO-6671 (P.I Rich), the data for the last two clusters (G302 and G312) are  
from the program GO-5906 (P.I. Holland) and were taken from the HST archive.      
\label{tab:log}}
\begin{tabular}{rrcccccl}  \\
\hline
\multicolumn{2}{c}{Cluster}&  V  &  $(B-V)$  &    R      &  R  &  Obs.  Date  \\
\multicolumn{1}{c}{Bo}&\multicolumn{1}{c}{G}&      &      &  (arcmin)  &(kpc)  &  \\
\hline
  293  &    11  &  16.30  &  0.74  &    75.75  &  16.95  &  09/25/1999  \\
  311  &    33  &  15.44  &  0.97  &    57.59  &  12.88  &  02/26/1999  \\
    12  &    64  &  15.08  &  0.77  &    25.35  &    5.67  &  08/17/1999  \\
  338  &    76  &  14.25  &  0.81  &    45.03  &  10.07  &  01/11/1999  \\
    27  &    87  &  15.60  &  0.92  &    26.43  &    5.91  &  08/16/1999  \\
    30  &    91  &  17.39  &  1.93  &    24.86  &    5.56  &  08/16/1999  \\
    58  &  119  &  14.97  &  0.84  &    30.59  &    6.84  &  06/13/1999  \\
  233  &  287  &  15.76  &  0.85  &    35.45  &    7.93  &  09/26/1999  \\
  384  &  319  &  15.75  &  0.99  &    72.07  &  16.12  &  02/28/1999  \\
  386  &  322  &  15.55  &  0.90  &    61.80  &  13.83  &  01/10/1999  \\
  240  &  302  &  15.21  &  0.72  &    31.77  &    7.11  &  11/05/1995  \\
  379  &  312  &  16.18  &  0.85  &    49.75  &  11.13  &  10/31/1995  \\
\hline  
\end{tabular}
\end{center}
Notes:  \\
1)  The integrated photometric parameters V and (B--V) are from Galleti et al.\ (2004). \\
2)  The  values  of  $R$  are  the  projected  galactocentric  distances  in  arcmin  
(Col.  5)  and  kpc  (Col.  6),  on  the  assumption  that  (m-M)$_0$=24.47$\pm$0.03  mag
(see Sect. 3.2.2).    
\end{table*}  

\clearpage  

\begin{table*}
\begin{center}
\caption{Information  on  the  sampled  population  for  each  cluster.  $Annulus$  represents  
the  total  area  where  photometry  was  performed.  N$_{before}$  and  N$_{after}$  are  the  
number  of  stars  before  and  after  field  subtraction.  $R>$  represents  the  radius  at  which  
photometric  blends  become  insignificant.  N$_{noblend}$  is  the  number  of  stars  
left  after  eliminating  the  photometric  blends.    }
\label{tab:fsub}  
\begin{tabular}{rccrrrr}\\
\hline
Cluster  &  $Annulus$  &  $L_{sam}/L_{tot}$  &  N$_{before}$  &  N$_{after}$  &  $R>$  &  N$_{noblend}$  \\
                &  (arcsec)    &                                      &                            &                          &(arcsec)&                          \\
\hline
G11          &  1.00  $-$10.35  &  0.56    &  805  &  722  &  2.76  &  411  \\
G33          &  1.70  $-$  5.98  &  0.31    &  697  &  468  &  3.22  &  256  \\
G64          &  1.84  $-$  7.82  &  0.27    &  687  &  558  &  3.22  &  287  \\
G76          &  2.76  $-$  7.28  &  0.17    &  960  &  776  &  3.68  &  463  \\
G87          &  1.56  $-$  5.52  &  0.32    &1123  &  836  &  2.30  &  610  \\
G91          &  1.50  $-$  3.50  &  0.21    &  193  &  133  &            &          \\
G119        &  2.48  $-$  5.52  &  0.12    &  745  &  484  &  3.22  &  386  \\
G287        &  1.56  $-$  4.74  &  0.29    &  756  &  587  &  2.76  &  219  \\
G319        &  1.61  $-$10.12  &  0.27    &  420  &  397  &  1.84  &  365  \\
G322        &  1.99  $-$  5.98  &  0.21    &  534  &  345  &  2.30  &  302  \\
G302        &  3.30  $-$12.00  &  0.21    &  712  &  409  &  6.00  &  229  \\
G312        &  2.50  $-$12.00  &  0.20    &  375  &  288  &  3.00  &  263  \\
\hline  
\end{tabular}
\end{center}
\end{table*}  

\clearpage  

\begin{table*}
\begin{center}
\caption{Reddening:  collection  of  values  from  the  literature,  and  adopted  
value.  }
\label{tab:ebmenv}
\scriptsize
\begin{tabular}{rcccccccccccc}  \\
\hline\hline
    &  &          &          &      &      &      &        &        &        &        &                  &          \\
Bo&G&Vet62&VdB69&FPC&BH  &SFD&C85a&C85b&B00a&B00b&adopted&$\sigma$  \\
  (1)  &(2)   &  (3)  &  (4)  &(5)&(6)&(7)&(8)  &(9)  &(10)  &(11)  &  (12) &  (13)  \\
\hline
293  &  11    &              &            &            &  0.08  &  0.06  &            &            &    0.12  &    0.12  &  0.09  &  0.03  \\
311  &  33    &    0.02  &  0.46  &  0.22  &  0.06  &  0.06  &  0.29  &  0.21  &    0.32  &    0.26  &  0.18  &  0.11  \\
  12  &  64    &  -0.09  &  0.21  &  0.17  &  0.04  &  0.06  &  0.05  &  0.07  &    0.14  &    0.10  &  0.09  &  0.05  \\
338  &  76    &    0.02  &  0.25  &            &  0.07  &  0.06  &            &            &    0.10  &    0.13  &  0.08  &  0.03  \\
  27  &  87    &    0.02  &  0.18  &  0.10  &  0.07  &  0.06  &  0.28  &  0.14  &    0.30  &    0.18  &  0.14  &  0.08  \\
  30  &  91    &    0.02  &  0.18  &  0.10  &  0.07  &  0.06  &  0.28  &  0.14  &    0.30  &    0.18  &  0.14  &  0.08  \\
  58  &  119  &    0.05  &  0.24  &  0.10  &  0.08  &  0.06  &            &            &    0.15  &    0.11  &  0.09  &  0.04  \\
233  &  287  &    0.09  &            &            &  0.07  &  0.06  &  0.08  &  0.07  &    0.18  &    0.14  &  0.09  &  0.05  \\
240  &  302  &  -0.01  &  0.21  &  0.10  &  0.09  &  0.06  &  0.06  &  0.13  &    0.08  &    0.06  &  0.08  &  0.02  \\
379  &  312  &              &            &            &  0.09  &  0.06  &            &            &    0.04  &    0.09  &  0.07  &  0.02  \\
384  &  319  &  -0.01  &            &            &  0.06  &  0.06  &            &            &    0.17  &    0.14  &  0.09  &  0.06  \\
386  &  322  &              &            &            &  0.18  &  0.06  &  0.27  &  0.10  &    0.17  &    0.02  &  0.13  &  0.06  \\
\hline
        &  1      &    0.16  &  0.19  &  0.10  &  0.05  &  0.06  &            &            &    0.01  &    0.08  &  0.07  &  0.02  \\
    6  &  58    &    0.07  &  0.08  &  0.10  &  0.04  &  0.06  &  0.18  &  0.11  &    0.16  &    0.13  &  0.10  &  0.05  \\
343  &  105  &  -0.09  &            &            &  0.05  &  0.06  &            &            &    0.09  &    0.05  &  0.06  &  0.01  \\
  45  &  108  &    0.23  &  0.31  &  0.10  &  0.03  &  0.06  &  0.12  &  0.16  &    0.17  &    0.17  &  0.10  &  0.06  \\
358  &  219  &  -0.16  &  0.15  &  0.10  &  0.04  &  0.06  &            &            &  -0.14  &  -0.04  &  0.06  &  0.08  \\
225  &  280  &    0.08  &  0.09  &  0.10  &  0.07  &  0.06  &            &            &    0.20  &    0.14  &  0.10  &  0.05  \\
405  &  351  &  -0.08  &  0.12  &            &  0.07  &  0.06  &            &            &    0.11  &    0.11  &  0.08  &  0.03  \\
468  &          &              &            &            &  0.04  &  0.06  &            &            &  -0.12  &  -0.29  &  0.06  &  0.15  \\

\hline\hline  
\end{tabular}
\end{center}
\scriptsize
Columns:  \\
(1): cluster identification from Battistini et al.\ (1987) \\
(2): cluster identification from Sargent et al.\ (1977) \\
(3):  from  Vetesnik (1962) \\
(4):  from  van den Bergh  (1969)  \\
(5):  from  Frogel et al.\ (1980)  \\
(6):  from  the HI maps (Burstein \& Heiles 1982) \\
(7):  from  the Galactic dust maps  (Schlegel et al.\ 1998) \\
(8):  from  E(B-V)=-0.066S+1.17(B-V)-0.32 (Crampton et al.\ 1985), (B-V) from Galleti 
et al.\ (2004) \\
(9):  from  E(B-V)=-0.066S+1.17(B-V)-0.32 (Crampton et al.\ 1985), (B-V) from  Crampton 
et al.\ (1985)\\
(10):  from  (B-V)$_0$=0.159[Fe/H]+0.92 (Barmby et al.\ 2000), [Fe/H] from Barmby et 
al.\ (2000), (B-V) from  Galleti et al.\ (2004) \\
(11):  from  (B-V)$_0$=0.159[Fe/H]+0.92 (Barmby et al.\ 2000), [Fe/H] and (B-V) from  
Barmby et al.\ (2000) \\
(12):  E(B-V)  final  adopted  value  \\
(13):  dispersion  of  the  mean  \\
\normalsize
\end{table*}  

\clearpage

\begin{table*}
\begin{center}
\caption[]{Observed  RGB  ridge  lines  (I,  V--I)  for  the  M31  GCs  considered  in  this  analysis.  
For  each  cluster  the  ridge  line  is  given  from  its  RGB  tip  in  steps  of  0.1  mag.
G91  is  missing  because  no  reliable  ridge  line  could  be  defined  (see  Sect. 3.2.1). }
\label{tab:ridge}
\scriptsize
\begin{tabular}{cccccccccc}
\hline\hline
              &  &    G11    &    G33    &    G64    &    G76    &    G87    &    G302  &    G312  &\\
\hline
        I    &  &  $(V-I)$  &  $(V-I)$  &  $(V-I)$  &  $(V-I)$  &  $(V-I)$  &  $(V-I)$  &  $(V-I)$  &\\  
\hline
  20.45  &  &              &              &              &  1.994  &  1.925  &              &              &\\
  20.55  &  &              &              &              &  1.928  &  1.879  &              &              &\\
  20.65  &  &  1.652  &              &              &  1.866  &  1.835  &              &              &\\
  20.75  &  &  1.616  &              &  1.494  &  1.808  &  1.794  &              &              &\\
  20.85  &  &  1.582  &              &  1.466  &  1.752  &  1.754  &              &              &\\
  20.95  &  &  1.546  &  1.721  &  1.440  &  1.700  &  1.717  &  1.591  &              &\\
  21.05  &  &  1.513  &  1.694  &  1.415  &  1.650  &  1.681  &  1.563  &              &\\
  21.15  &  &  1.482  &  1.665  &  1.392  &  1.603  &  1.647  &  1.535  &              &\\
  21.25  &  &  1.454  &  1.639  &  1.369  &  1.560  &  1.615  &  1.507  &              &\\
  21.35  &  &  1.428  &  1.612  &  1.348  &  1.518  &  1.585  &  1.479  &              &\\
  21.45  &  &  1.404  &  1.589  &  1.328  &  1.479  &  1.557  &  1.453  &              &\\
  21.55  &  &  1.383  &  1.567  &  1.309  &  1.443  &  1.529  &  1.427  &  1.920  &\\
  21.65  &  &  1.362  &  1.548  &  1.292  &  1.410  &  1.504  &  1.404  &  1.842  &\\
  21.75  &  &  1.343  &  1.529  &  1.275  &  1.378  &  1.479  &  1.384  &  1.772  &\\
  21.85  &  &  1.325  &  1.511  &  1.260  &  1.348  &  1.456  &  1.367  &  1.716  &\\
  21.95  &  &  1.307  &  1.494  &  1.245  &  1.321  &  1.435  &  1.350  &  1.673  &\\
  22.05  &  &  1.290  &  1.480  &  1.232  &  1.296  &  1.415  &  1.334  &  1.638  &\\
  22.15  &  &  1.274  &  1.465  &  1.220  &  1.273  &  1.397  &  1.319  &  1.608  &\\
  22.25  &  &  1.258  &  1.452  &  1.209  &  1.251  &  1.379  &  1.305  &  1.577  &\\
  22.35  &  &  1.242  &  1.439  &  1.197  &  1.230  &  1.362  &  1.292  &  1.548  &\\
  22.45  &  &  1.228  &  1.427  &  1.186  &  1.211  &  1.347  &  1.279  &  1.522  &\\
  22.55  &  &  1.214  &  1.415  &  1.175  &  1.194  &  1.332  &  1.267  &  1.498  &\\
  22.65  &  &  1.202  &  1.404  &  1.164  &  1.178  &  1.318  &  1.256  &  1.477  &\\
  22.75  &  &  1.190  &  1.392  &  1.153  &  1.163  &  1.305  &  1.244  &  1.455  &\\
  22.85  &  &  1.179  &  1.381  &  1.143  &  1.149  &  1.293  &  1.234  &  1.432  &\\
  22.95  &  &  1.168  &  1.370  &  1.134  &  1.137  &  1.282  &  1.224  &  1.410  &\\
  23.05  &  &  1.158  &  1.358  &  1.124  &  1.126  &  1.271  &  1.214  &  1.389  &\\
  23.15  &  &  1.149  &  1.347  &  1.115  &  1.114  &  1.262  &  1.205  &  1.370  &\\
  23.25  &  &  1.139  &  1.336  &  1.107  &  1.104  &  1.252  &  1.196  &  1.352  &\\
  23.35  &  &  1.131  &  1.325  &  1.098  &  1.094  &  1.243  &  1.187  &  1.336  &\\
  23.45  &  &  1.122  &  1.315  &  1.090  &  1.085  &  1.235  &  1.178  &  1.319  &\\
  23.55  &  &  1.114  &  1.306  &  1.082  &  1.076  &  1.226  &  1.170  &  1.302  &\\
  23.65  &  &  1.106  &  1.296  &  1.074  &  1.067  &  1.218  &  1.161  &  1.284  &\\
  23.75  &  &  1.098  &  1.287  &  1.067  &  1.058  &  1.211  &  1.153  &  1.267  &\\
  23.85  &  &  1.090  &  1.279  &  1.060  &  1.050  &  1.203  &  1.144  &  1.250  &\\
  23.95  &  &  1.083  &  1.272  &  1.052  &  1.041  &  1.195  &  1.136  &  1.235  &\\
  24.05  &  &  1.077  &  1.264  &  1.046  &  1.034  &  1.188  &  1.127  &  1.221  &\\
  24.15  &  &  1.071  &  1.257  &  1.039  &  1.026  &  1.181  &  1.119  &  1.208  &\\
  24.25  &  &  1.065  &  1.250  &  1.033  &  1.017  &  1.173  &  1.110  &  1.197  &\\
  24.35  &  &  1.059  &  1.243  &  1.027  &  1.006  &  1.165  &  1.102  &  1.187  &\\
  24.45  &  &  1.053  &  1.236  &  1.021  &              &              &  1.094  &  1.178  &\\
  24.55  &  &  1.046  &  1.230  &  1.015  &              &              &              &  1.169  &\\
  24.65  &  &  1.039  &  1.223  &  1.010  &              &              &              &  1.162  &\\
  24.75  &  &  1.032  &  1.216  &  1.004  &              &              &              &  1.154  &\\
  24.85  &  &              &  1.210  &  0.999  &              &              &              &  1.147  &\\
  24.95  &  &              &  1.203  &  0.994  &              &              &              &              &\\
  25.05  &  &              &              &  0.989  &              &              &              &              &\\
  25.15  &  &              &              &  0.985  &              &              &              &              &\\
  25.25  &  &              &              &  0.981  &              &              &              &              &\\
  &  &  &  &  &  &  &  &  &\\    
\end{tabular}    
\end{center}
\normalsize
\end{table*}
\begin{table*}
Table  4:  -  continued  -\\
\scriptsize
\begin{center}
\begin{tabular}{ccccccc}
\hline
              &  &    G119  &    G287  &    G319  &    G322  &\\  
\hline
      I      &  &  $(V-I)$  &  $(V-I)$  &  $(V-I)$  &  $(V-I)$  &  \\  
\hline
  20.80  &  &              &              &  2.305  &  1.725  &\\
  20.90  &  &              &              &  2.196  &  1.680  &\\
  21.00  &  &              &  1.538  &  2.086  &  1.639  &\\
  21.10  &  &  1.564  &  1.493  &  1.982  &  1.606  &\\
  21.20  &  &  1.538  &  1.457  &  1.893  &  1.573  &\\
  21.30  &  &  1.513  &  1.429  &  1.827  &  1.539  &\\
  21.40  &  &  1.490  &  1.399  &  1.775  &  1.509  &\\
  21.50  &  &  1.467  &  1.372  &  1.725  &  1.478  &\\
  21.60  &  &  1.444  &  1.347  &  1.676  &  1.449  &\\
  21.70  &  &  1.422  &  1.322  &  1.629  &  1.422  &\\
  21.80  &  &  1.400  &  1.300  &  1.586  &  1.396  &\\
  21.90  &  &  1.379  &  1.281  &  1.545  &  1.372  &\\
  22.00  &  &  1.360  &  1.263  &  1.511  &  1.348  &\\
  22.10  &  &  1.343  &  1.243  &  1.475  &  1.326  &\\
  22.20  &  &  1.327  &  1.223  &  1.441  &  1.304  &\\
  22.30  &  &  1.312  &  1.207  &  1.409  &  1.282  &\\
  22.40  &  &  1.298  &  1.192  &  1.379  &  1.261  &\\
  22.50  &  &  1.286  &  1.179  &  1.350  &  1.240  &\\
  22.60  &  &  1.273  &  1.166  &  1.322  &  1.220  &\\
  22.70  &  &  1.261  &  1.154  &  1.294  &  1.200  &\\
  22.80  &  &  1.249  &  1.142  &  1.269  &  1.181  &\\
  22.90  &  &  1.238  &  1.130  &  1.245  &  1.162  &\\
  23.00  &  &  1.227  &  1.119  &  1.222  &  1.144  &\\
  23.10  &  &  1.216  &  1.106  &  1.200  &  1.127  &\\
  23.20  &  &  1.206  &  1.095  &  1.179  &  1.110  &\\
  23.30  &  &  1.196  &  1.084  &  1.160  &  1.094  &\\
  23.40  &  &  1.186  &  1.074  &  1.142  &  1.080  &\\
  23.50  &  &  1.176  &  1.065  &  1.124  &  1.065  &\\
  23.60  &  &  1.166  &  1.057  &  1.109  &  1.052  &\\
  23.70  &  &  1.156  &  1.050  &  1.096  &  1.038  &\\
  23.80  &  &  1.145  &  1.043  &  1.084  &  1.026  &\\
  23.90  &  &  1.135  &  1.036  &  1.073  &  1.015  &\\
  24.00  &  &  1.124  &  1.029  &  1.063  &  1.006  &\\
  24.10  &  &  1.114  &  1.022  &  1.052  &  0.998  &\\
  24.20  &  &  1.104  &  1.014  &  1.042  &  0.990  &\\
  24.30  &  &  1.094  &  1.007  &  1.033  &  0.982  &\\
  24.40  &  &  1.085  &  1.000  &  1.025  &  0.974  &\\
  24.50  &  &  1.076  &  0.992  &  1.017  &  0.967  &\\
  24.60  &  &  1.068  &  0.985  &  1.009  &  0.959  &\\
  24.70  &  &  1.060  &  0.979  &  1.002  &  0.951  &\\
  24.80  &  &  1.052  &  0.972  &  0.995  &  0.944  &\\
  24.90  &  &  1.045  &  0.967  &  0.989  &  0.938  &\\
  25.00  &  &  1.037  &  0.962  &  0.982  &  0.932  &\\
  25.10  &  &  1.030  &  0.957  &  0.976  &  0.926  &\\
  25.20  &  &  1.022  &              &  0.969  &  0.922  &\\
  25.30  &  &  1.015  &              &  0.962  &  0.918  &\\
  25.40  &  &              &              &  0.956  &              &\\
  25.50  &  &              &              &              &              &\\
  25.60  &  &              &              &              &              &\\
  25.70  &  &              &              &              &              &\\
  25.80  &  &              &              &              &              &\\
  &  &  &  &  &  &\\
\end{tabular}
\end{center}
\normalsize
\end{table*}

\clearpage

\begin{table*}
\begin{center}
\caption{Estimates  of  metallicity  [Fe/H]  for  the    M31  GCs  considered  in  this  
analysis  (upper  group  of  12  clusters)  and  those  studied  in  Paper  I  (lower  
group  of  8  clusters).    The  metallicity  in  the  ZW84  and  CG97  metallicity  
scales  were  calculated  according  to  Saviane  et  al.\  (2000,  Table  6).  
For  G280  and  G351  metallicities  could  only  be  derived  in  the  CG97  scale  
(see  Ferraro  et  al.\  1999,  Table  4).  
}
\label{tab:mets}  
\begin{tabular}{rr|ccc|cc|cc|c|c  c}  \\
\hline\hline
Bo  &  G  &  S  &  RGB$_{ZW}$  &  RGB$_{CG}$  &  Ridge    &  Ridge  &  HBK  &  P  &  B  &  Adopted  &  $\sigma$\\
      &      &(1)&  (2)                &  (3)                &  (4)        &  (5)      &  (6)  &(7)&(8)  &  (9)  &  (10)\\
\hline
293  &  11    &  8.23    &  -1.70  &  -1.33  &  -1.6  &  -1.7      &-1.89  &              &  -2.13  &  -1.80  &  0.21\\
311  &  33    &  8.47    &  -1.75  &  -1.39  &  -1.6  &  -1.75    &-1.74  &  -1.96  &  -1.88  &  -1.78  &  0.13\\
  12  &  64    &  8.45    &  -1.75  &  -1.39  &  -1.8  &  -1.9      &-1.81  &  -1.65  &  -2.17  &  -1.85  &  0.18\\
338  &  76    &  4.67    &  -0.84  &  -0.77  &  -1.3  &  -1.3      &-1.34  &  -1.46  &  -1.41  &  -1.28  &  0.22\\
  27  &  87    &  8.37    &  -1.79  &  -1.37  &  -1.4  &  -1.66    &-1.64  &              &  -1.72  &  -1.64  &  0.15\\
  30  &  91    &      &            &                        &            &                &            &  -0.39  &              &  \\
  58  &  119  &  7.06    &  -1.41  &  -1.06  &  -1.3  &  -1.4      &-1.45  &              &  -1.35  &  -1.38  &  0.06\\
233  &  287  &  7.81    &  -1.60  &  -1.22  &  -1.6  &  -1.6      &-1.59  &              &              &  -1.60  &  0.01\\
240  &  302  &  7.91    &  -1.62  &  -1.25  &  -1.4  &  -1.66    &-1.76  &              &  -1.97  &  -1.68  &  0.21\\
379  &  312  &  3.25    &  -0.50  &  -0.76  &  -0.51&  -0.6      &-0.70  &              &  -0.88  &  -0.64  &  0.16\\
384  &  319  &  3.87    &  -0.65  &  -0.75  &  -0.7  &  -0.7      &-0.66  &              &  -1.62  &  -0.87  &  0.42\\
386  &  322  &  4.62    &  -0.83  &  -0.77  &    np    &  -1.2      &-1.21  &  -1.62  &  -0.57  &  -1.09  &  0.40\\
\hline        \hline
        &  1      &  4.21    &  -0.73  &  -0.75  &  -1.0  &  -0.9      &-1.08  &              &  -1.23  &  -0.99  &  0.19\\
    6  &  58    &  3.25    &  -0.50  &  -0.76  &  -0.6  &  -0.6      &-0.57  &  -0.58  &  -0.59  &  -0.57  &  0.04\\
343  &  105  &  7.34    &  -1.48  &  -1.12  &  -1.45&  -1.5      &-1.49  &              &  -1.29  &  -1.44  &  0.09\\
  45  &  108  &  4.78    &  -0.87  &  -0.78  &  -0.6  &  -0.9      &-0.94  &  -1.05  &  -0.76  &  -0.85  &  0.16\\
358  &  219  &  8.85    &  -1.84  &  -1.50  &  -1.91&  -1.91    &-1.83  &              &  -2.09  &  -1.92  &  0.10\\
225  &  280  &  3.82    &              &  -0.40  &  -0.35&  -0.71    &-0.70  &  -0.67  &  -0.37  &  -0.56  &  0.18\\
405  &  351  &  5.54    &              &  -0.88  &  -1.65&  -1.70    &-1.80  &              &  -1.93  &  -1.77  &  0.12\\
468  &          &  4.23    &  -0.74  &  -0.75  &  -0.7  &  -0.75    &            &              &              &  -0.73  &  0.03\\
\hline\hline  
\end{tabular}
\end{center}
Notes:  \\
(1):  The  $S$  parameter  is  defined  in  the  (V,V--I)  plane,  except  for  clusters  
G280  and  G351  where  the  $S_{2.0}$  parameter,  defined  in  the  (V,B--V)  plane,  
is  reported  instead  (Saviane  et  al.\  2000);  \\
(2):  from  the  $S$-parameter,  ZW84  metallicity  scale;  \\
(3):  from  the  $S$-parameter,  CG97  metallicity  scale;  \\
(4):  from  the  comparison  of  the  RGB  ridge  lines  with  GGC  templates,  using  the  
adopted  E(B-V);  \\
(5):  from  the  comparison  of  the  RGB  ridge  lines  with  GGC  templates,  using  
E(B-V)  as  a  free  parameter;  \\
(6):  spectroscopic  (Huchra et al.\ 1991; Barmby et al.\ 2000) \\
(7):  spectroscopic  (Perrett et al.\ 2002)  \\
(8):  photometric  (V--K  integrated  colors, Bonoli et al.\ 1987) \\
(9):  Adopted  values  (see  Sect.  3.2.3)  \\
(10):  Dispersion  of  the  mean  (see  Sect.  3.2.3)  \\
\end{table*}  

\clearpage  

\begin{table*}
\begin{center}
\caption{Reference  grid  of  template  Galactic  GCs.  Metallicities  are  from  
Zinn (1985); all  other  parameters  are  from  Harris  (1996,  update  2003).
}
\label{tab:grid}
\begin{tabular}{lccc}  \\
\hline
Cluster  &  ${\rm[Fe/H]}_{ZW84}$  &  $E(B-V)$  &  $(m-M)_V$    \\
\hline
M15          &  -2.15  &    0.10  &  15.37  \\
NGC6397  &  -1.91  &    0.18  &  12.36  \\
NGC5824  &  -1.87  &    0.13  &  17.93  \\
M13          &  -1.65  &    0.02  &  14.48  \\
M3            &  -1.66  &    0.01  &  15.12  \\
M5            &  -1.40  &    0.03  &  14.46  \\
47  Tuc    &  -0.71  &    0.04  &  13.37  \\
NGC6356  &  -0.62  &    0.28  &  16.77  \\
NGC6624  &  -0.35  &    0.28  &  15.36  \\
NGC6553  &  -0.29  &    0.63  &  15.83  \\
\hline  
\end{tabular}
\end{center}
\end{table*}  

\clearpage  

\begin{table*}
\begin{center}
\caption{Metallicities  and  individual  distance  moduli  using  adopted  and  fitted
reddening. The adopted values are described in Sect. 3.1 and are listed in 
Table \ref{tab:ebmenv}. The fitted values are described in Sect. 3.2.2.}
\label{tab:ebvmod}
\begin{tabular}{rr|ccc|ccc}  \\
\hline\hline
Bo  &  G      &  E(B-V)$_{ad}$  &[Fe/H]  &(m-M)$_{0}$&E(B-V)$_{free}$  &[Fe/H]  &(m-M)$_{0}$      \\

\hline  
293  &  11    &  0.09    &  -1.6  &  24.52    &  0.11    &  -1.7    &  24.48      \\
311  &  33    &  0.18    &  -1.6  &  24.53    &  0.25    &  -1.75  &  24.30      \\
  12  &  64    &  0.09    &  -1.8  &  24.40    &  0.11    &  -1.9    &  24.30      \\
338  &  76    &  0.08    &  -1.3  &  24.20    &  0.05    &  -1.3    &  24.35      \\
  27  &  87    &  0.14    &  -1.4  &  24.55    &  0.18    &  -1.66  &  24.55      \\
  58  &  119  &  0.09    &  -1.3  &  24.35    &  0.12    &  -1.4    &  24.35      \\
233  &  287  &  0.09    &  -1.6  &  24.36    &  0.08    &  -1.6    &  24.42      \\
240  &  302  &  0.08    &  -1.4  &  24.60    &  0.14    &  -1.66  &  24.40      \\
379  &  312  &  0.07    &  -0.51&  24.53    &  0.11    &  -0.6    &  24.44      \\
384  &  319  &  0.09    &  -0.7  &  24.43    &  0.04    &  -0.7    &  24.55      \\
386  &  322  &  0.13    &    np    &  24.30    &  0.04    &  -1.2    &  24.50      \\
\hline
        &  1      &  0.07    &  -1.0  &  24.46    &  0.04    &  -0.9    &  24.55      \\
    6  &  58    &  0.10    &  -0.6  &  24.57    &  0.09    &  -0.6    &  24.56      \\
343  &  105  &  0.06    &  -1.45&  24.68    &  0.07    &  -1.5    &  24.68      \\
  45  &  108  &  0.10    &  -0.6  &  24.55    &  0.15    &  -0.9    &  24.55      \\
358  &  219  &  0.06    &  -1.91&  24.52    &  0.05    &  -1.91  &  24.58      \\
225  &  280  &  0.10    &  -0.35&  24.53    &  0.15    &  -0.71  &  24.40      \\
405  &  351  &  0.08    &  -1.65&  24.73    &  0.09    &  -1.70  &  24.68      \\
468  &          &  0.06    &  -0.7  &  24.48    &  0.08    &  -0.75  &  24.50      \\
\hline\hline  
\end{tabular}
\end{center}
\end{table*}  

\clearpage

\begin{table*}
\begin{center}
\caption{Adopted  reddenings,  metallicities  and  distances  from  the  RGB  
morphology  (see  Sect.  3.1.2).  The  Mironov  index  B/(B+R)  is  discussed  in  
Sect.  3.2.1;  the  V(HB)  magnitudes  and the distance  moduli are  discussed  in  
Sect.  3.2.2.}
\label{tab:vhb}
\scriptsize
\begin{tabular}{rr|ccc||cc|cc|cc|cc|c}  \\
\hline
\multicolumn{2}{c}{Cluster}  &  $E(B-V)$  &  [Fe/H]  &  V(HB)  &  M$_{V}$&$(m-M)_0$  &M$_{V}$&  $(m-M)_0$  &
M$_{V}$&  $(m-M)_0$  &M$_{V}$&  $(m-M)_0$  &  B/(B+R)  \\
\multicolumn{1}{c}{Bo}&\multicolumn{1}{c}{G}&  &    &    &\multicolumn{2}{c}{(A)}&\multicolumn{2}{c}{(B)}&
\multicolumn{2}{c}{(C)}&\multicolumn{2}{c}{(D)}&  \\

\hline      
293  &  11    &    0.09  &-1.80  &  25.23  &  0.43  &  24.52  &  0.41  &  24.48  &  0.48  &  24.47  &  0.48  &  24.47  &  0.84  \\
311  &  33    &    0.18  &-1.78  &  25.49  &  0.40  &  24.53  &  0.41  &  24.30  &  0.49  &  24.45  &  0.46  &  24.47  &  0.44  \\
 12  &  64    &    0.09  &-1.85  &  24.92  &  0.24  &  24.40  &  0.28  &  24.30  &  0.47  &  24.17  &  0.17  &  24.47  &  0.69  \\
338  &  76    &    0.08  &-1.28  &  24.94  &  0.49  &  24.20  &  0.43  &  24.35  &  0.60  &  24.09  &  0.22  &  24.47  &  0.43  \\
 27  &  87    &    0.14  &-1.64  &  25.51  &  0.53  &  24.55  &  0.40  &  24.55  &  0.52  &  24.56  &  0.61  &  24.47  &  0.17  \\
 58  &  119   &    0.09  &-1.38  &  25.21  &  0.58  &  24.35  &  0.49  &  24.35  &  0.58  &  24.35  &  0.46  &  24.47  &  0.27  \\
233  &  287   &    0.09  &-1.60  &  25.24  &  0.60  &  24.36  &  0.57  &  24.42  &  0.53  &  24.43  &  0.49  &  24.47  &  0.35  \\
240  &  302   &    0.08  &-1.68  &  25.14  &  0.29  &  24.60  &  0.31  &  24.40  &  0.51  &  24.38  &  0.42  &  24.47  &  0.53  \\
379  &  312   &    0.07  &-0.64  &  25.50  &  0.75  &  24.53  &  0.72  &  24.44  &  0.74  &  24.55  &  0.81  &  24.47  &  0.00  \\
384  &  319   &    0.09  &-0.87  &  25.46  &  0.75  &  24.43  &  0.79  &  24.55  &  0.69  &  24.49  &  0.71  &  24.47  &  0.00  \\
386  &  322   &    0.13  &-1.09  &  25.10  &  0.40  &  24.30  &  0.48  &  24.50  &  0.64  &  24.06  &  0.23  &  24.47  &  0.41  \\
\hline  
     &  1     &    0.07  &-0.99  &  25.15  &  0.47  &  24.46  &  0.48  &  24.55  &  0.66  &  24.27  &  0.46  &  24.47  &  0.19  \\
  6  &  58    &    0.10  &-0.57  &  25.46  &  0.58  &  24.57  &  0.62  &  24.56  &  0.75  &  24.40  &  0.68  &  24.47  &  0.03  \\
343  &  105   &    0.06  &-1.44  &  25.47  &  0.60  &  24.68  &  0.57  &  24.68  &  0.56  &  24.73  &  0.81  &  24.47  &  0.74  \\
 45  &  108   &    0.10  &-0.85  &  25.62  &  0.76  &  24.55  &  0.61  &  24.55  &  0.69  &  24.62  &  0.84  &  24.47  &  0.14  \\
358  &  219   &    0.06  &-1.92  &  25.27  &  0.56  &  24.52  &  0.54  &  24.58  &  0.46  &  24.63  &  0.61  &  24.47  &  0.78  \\
225  &  280   &    0.10  &-0.56  &  25.52  &  0.68  &  24.53  &  0.66  &  24.40  &  0.76  &  24.45  &  0.74  &  24.47  &  0.15  \\
405  &  351   &    0.08  &-1.77  &  25.43  &  0.45  &  24.73  &  0.47  &  24.68  &  0.49  &  24.69  &  0.71  &  24.47  &  0.71  \\
468  &        &    0.06  &-0.73  &  25.41  &  0.74  &  24.48  &  0.66  &  24.50  &  0.72  &  24.51  &  0.75  &  24.47  &  0.00  \\
\hline
\end{tabular}
\end{center}
Notes:  \\
(A): adopted reddening (column 2 this table), best fit distance, derived $M_V$; \\
(B): best fit reddening and distance, derived $M_V$;  \\ 
(C): $M_V$ from eq. (1), adopted reddening (column 2 this table), derived distance; \\
(D): adopted distance (24.47) and reddening (column 2 this table), derived $M_V$ and hence eq.
(2). \\
\normalsize
\end{table*}  

\clearpage



\begin{figure}
\epsscale{0.8}
\caption{I band (F814W) images of the 12 M31 GCs analysed in the present study. 
All of the $20^{\prime\prime}\times 20^{\prime\prime} (=80 pc)$ subrasters are from PC 
frames except for those of G91, G302 and G312 which fell on the WFC.  
The clusters are displayed in the order of Fig. \ref{fig:cmd}.
Available in $jpeg$ format. 
}
\label{fig:images}
\end{figure}

\clearpage

\begin{figure}
\epsscale{0.8}
\plotone{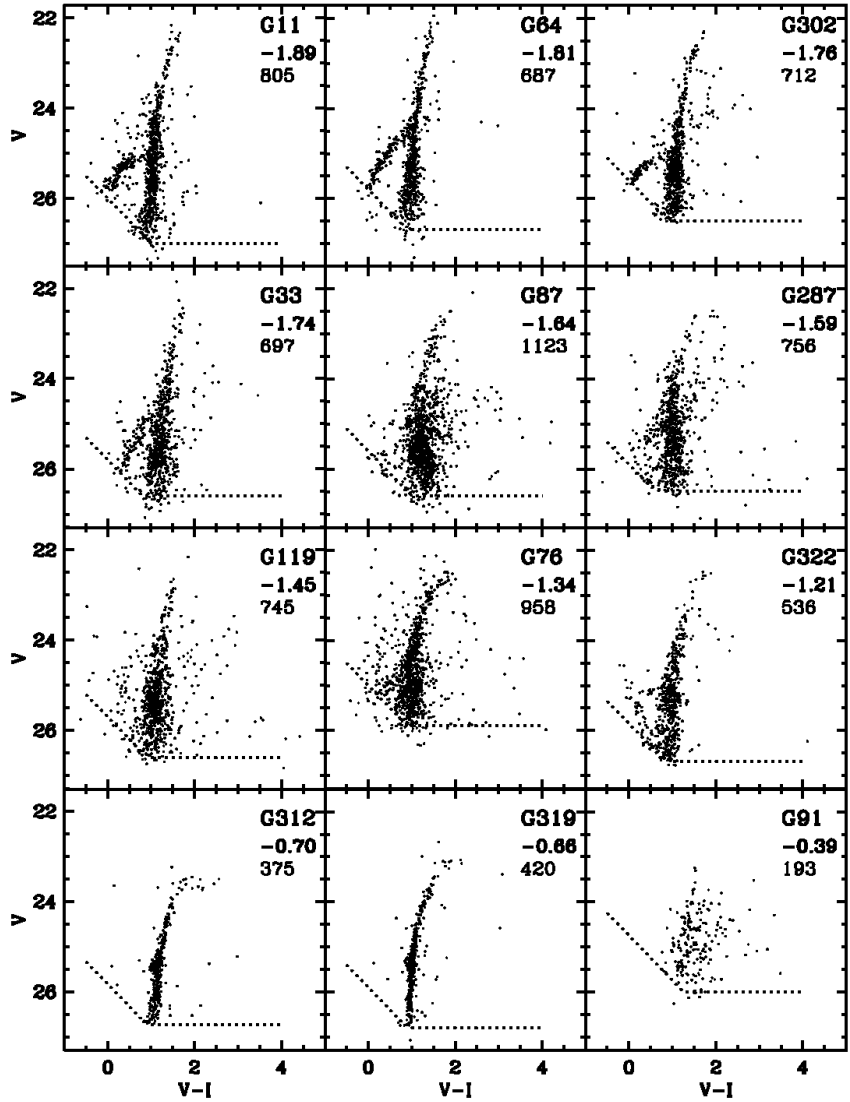}
\caption{Observed  CMDs  for  the  12  M31  GCs  analysed  in  the  present  study.  
Each  cluster  is  labelled  with  its  name,  metallicity,  and  number  of  stars  
displayed  (i.e.  all  the  stars  detected  and  measured  within  the  annuli  listed  
in  Table  \ref{tab:fsub}).  The  dotted  lines  represent  the  limiting  magnitude  
cutoffs  of  the  photometry.  
}
\label{fig:cmd}
\end{figure}

\clearpage

\begin{figure}
\plotone{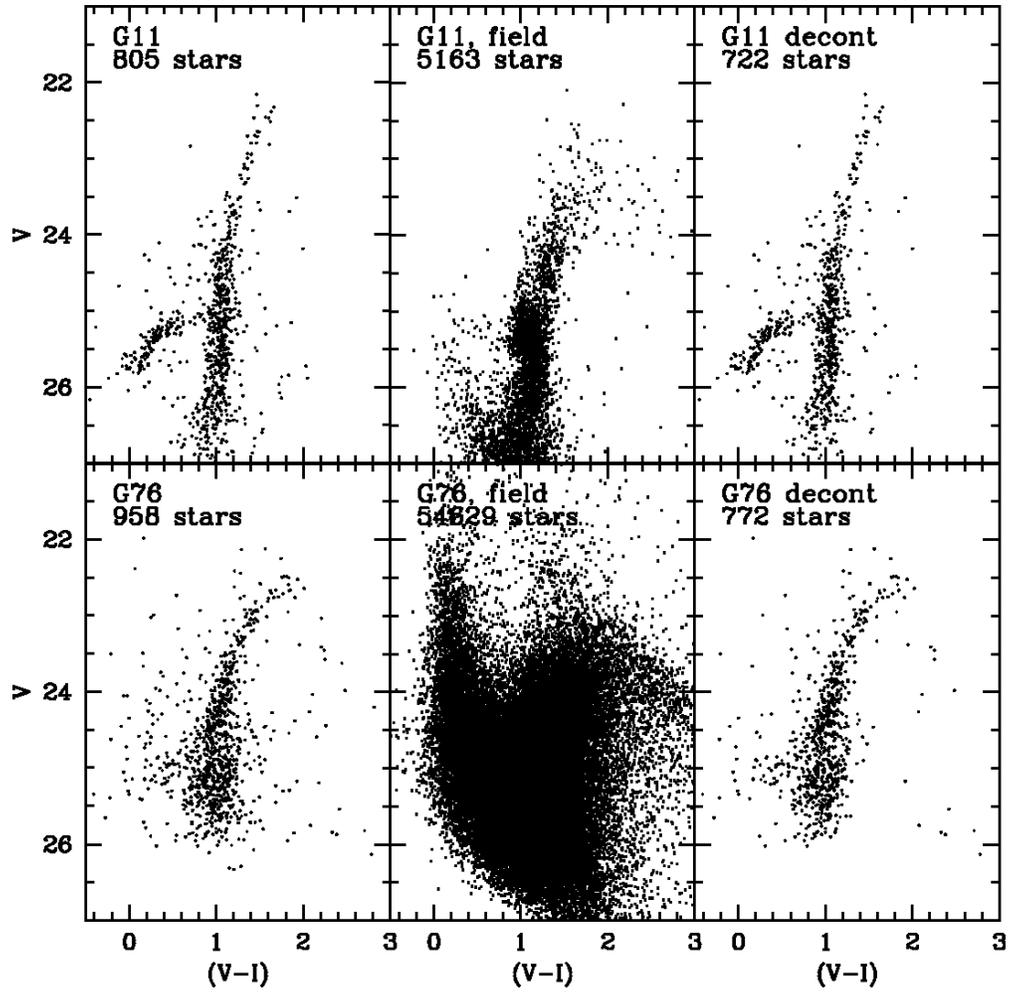}
\caption{Examples  of  statistical  field  subtraction:  CMD  of  G76  and  G11,  
before  (left)  and  after  (right)  field  subtraction.  The  middle  panel  shows  
the  CMD  of  the  corresponding  field,  defined  over  a  much  larger  WFC  area.  
}
\label{fig:fieldsub}
\end{figure}

\clearpage

\begin{figure}
\plotone{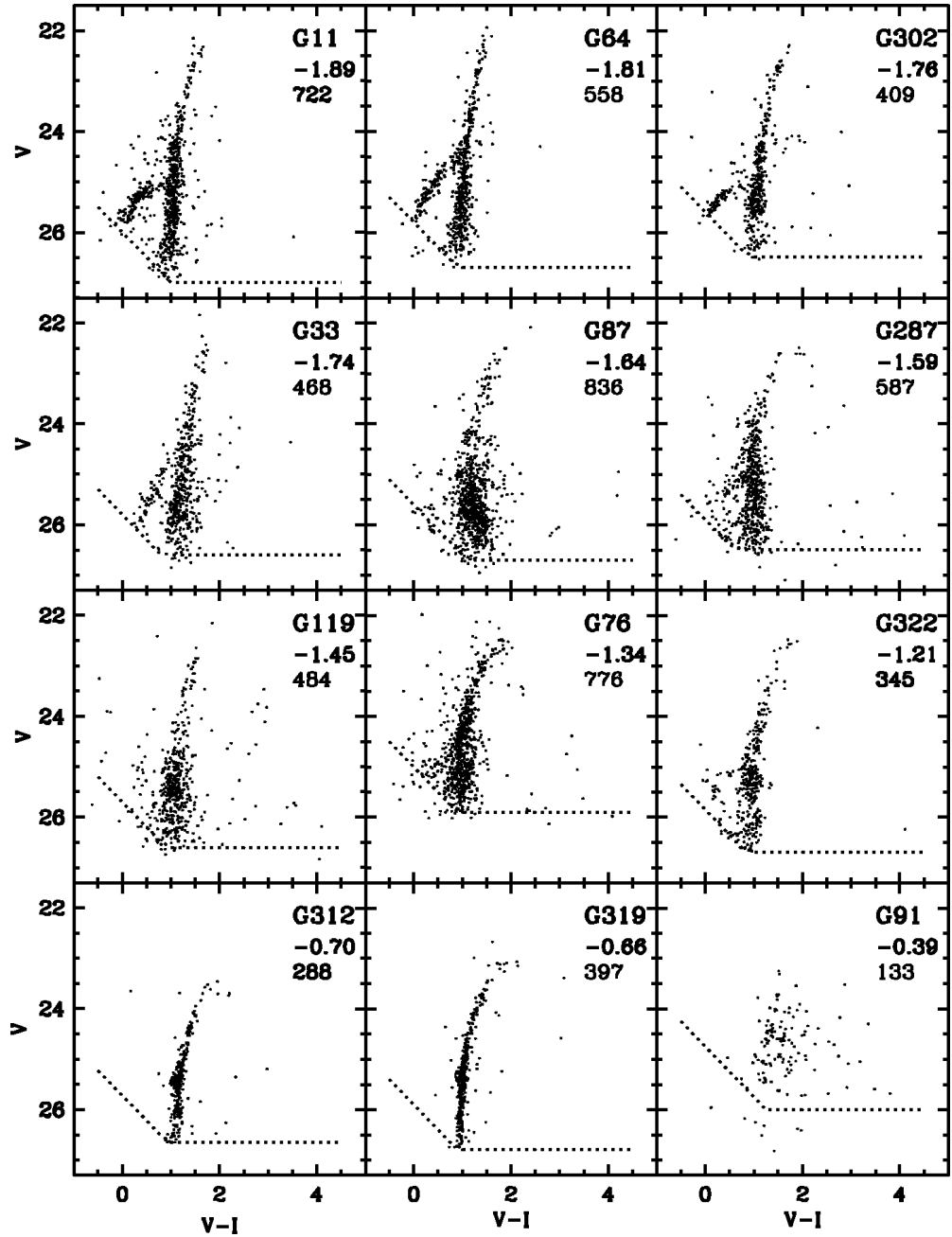}
\caption{As in  Fig.  \ref{fig:cmd},  where  the  CMDs  have  been  cleaned  of  field  
contamination.  The  number  of  stars  displayed  in  each  panel  represents  those  
that  are  left  after  field  subtraction.  
}
\label{fig:cmdsub}
\end{figure}

\clearpage

\begin{figure}
\epsscale{0.75}
\plotone{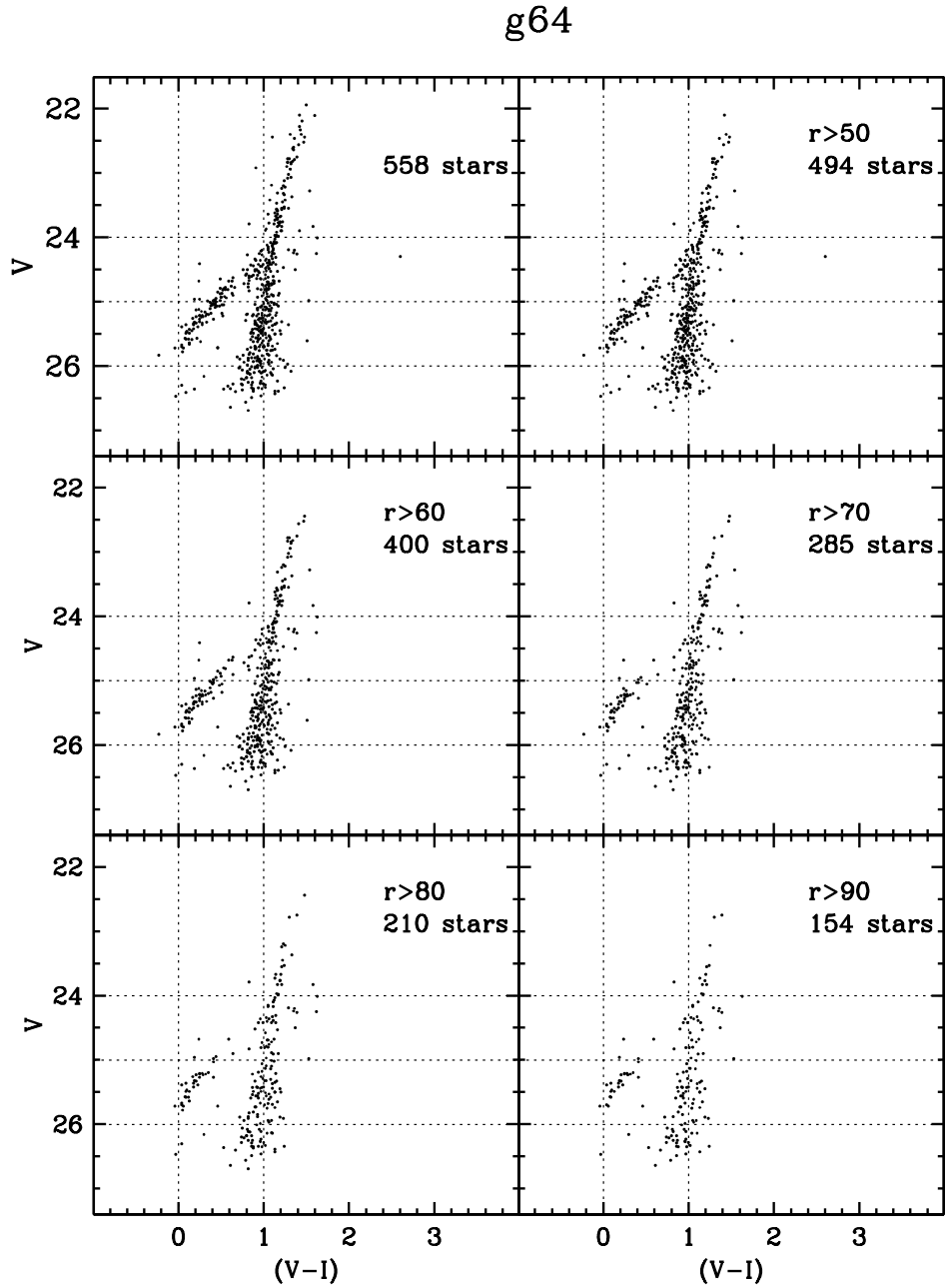}
\caption{CMD  of  G64:  the  top  left  panel  shows  the  entire  photometric  
data  set,  the  other  panels  show  the  stars  measured  in  progressively  
external  areas  of  radius  $r$  (in  pixels).  The  brightest  and  reddest  
part  of  what  might  look  like  the  HB  in  the  total  CMD  progressively  
disappears  in  the  CMDs  of  more  external  less  crowded  areas,  thus  
indicating  that  these  are  not  true  stars  but  photometric  blends.  
The  radius  at  which  the  blending  becomes  irrelevant  is  at  
$r\sim$70  px.    
}
\label{fig:g64blend}
\end{figure}

\clearpage

\begin{figure}
\epsscale{0.8}
\plotone{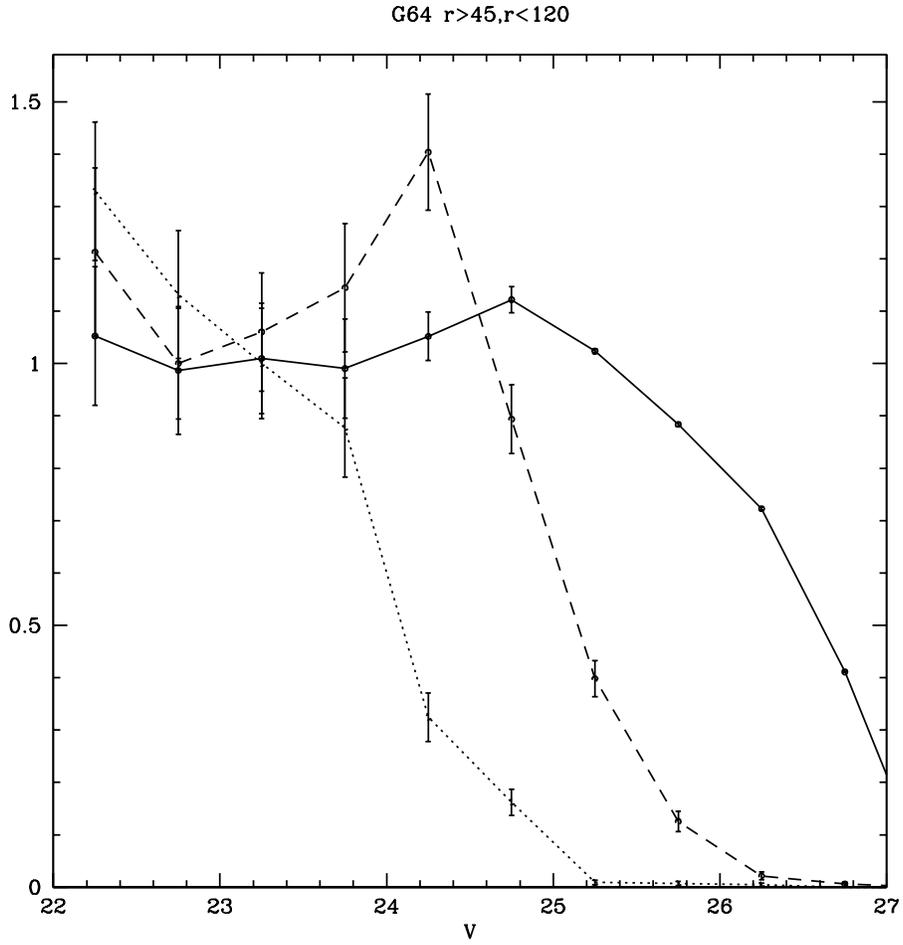}
\caption{Completeness curves for G64 over different radial annuli. Dotted:
45 $< r <$ 60 px; dashed: 60 $< r <$ 80 px; full: $r>$ 80 px. 
See sect. 2.5 for discussion.  
}
\label{fig:complete}
\end{figure}

\clearpage

\begin{figure}
\epsscale{0.8}
\plotone{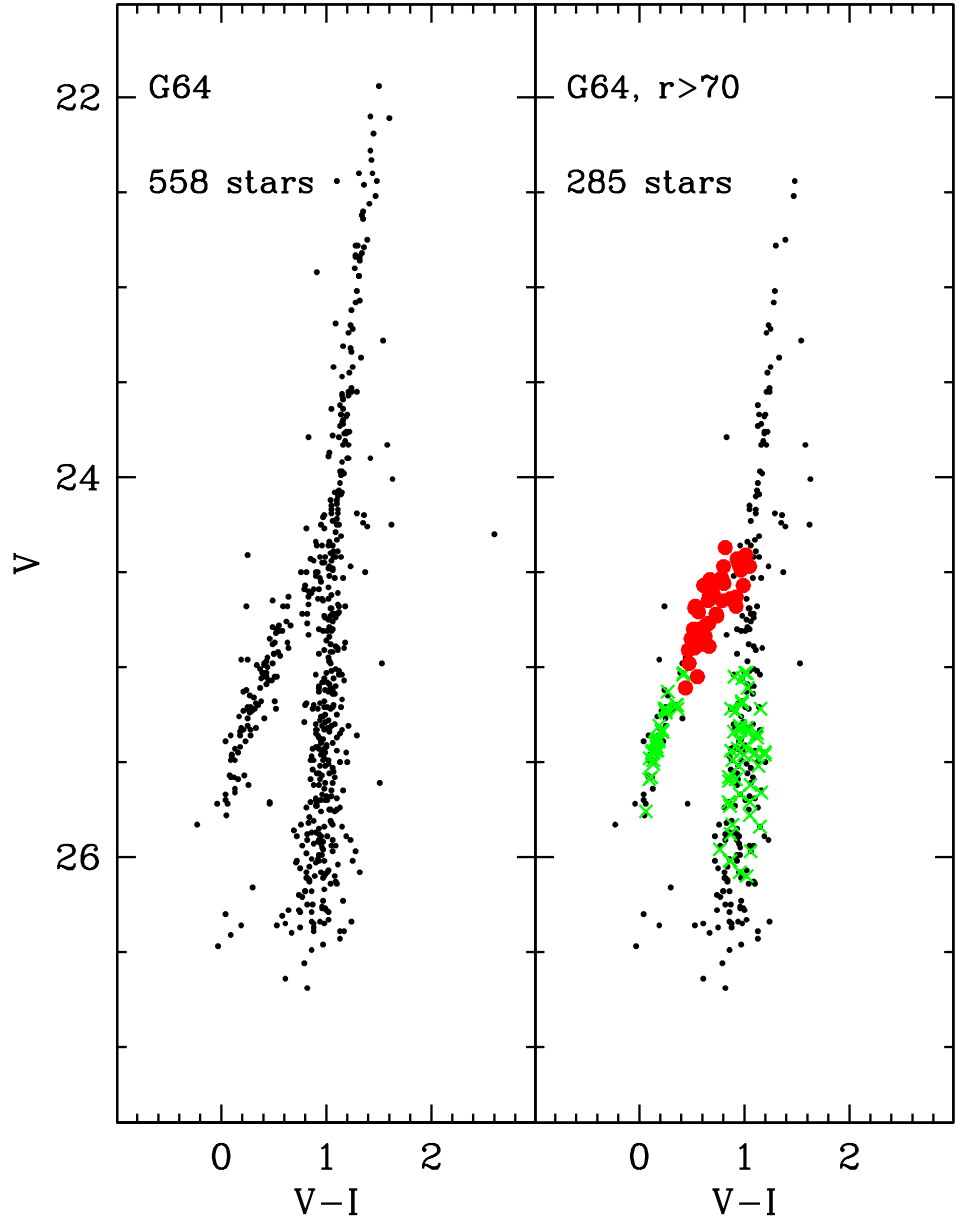}
\caption{CMDs for G64: the left panel shows the entire CMD after the subtraction  
of  the  field;  in  the right panel we have superposed to a more external CMD 
(r$ > $70 px) several 'fake' simulated  stars (shown as red dots) resulting from 
the  photometric  blend  of HB and RGB stars (shown as green crosses).  
}
\label{fig:simg64}
\end{figure}

\clearpage

\begin{figure}
\plotone{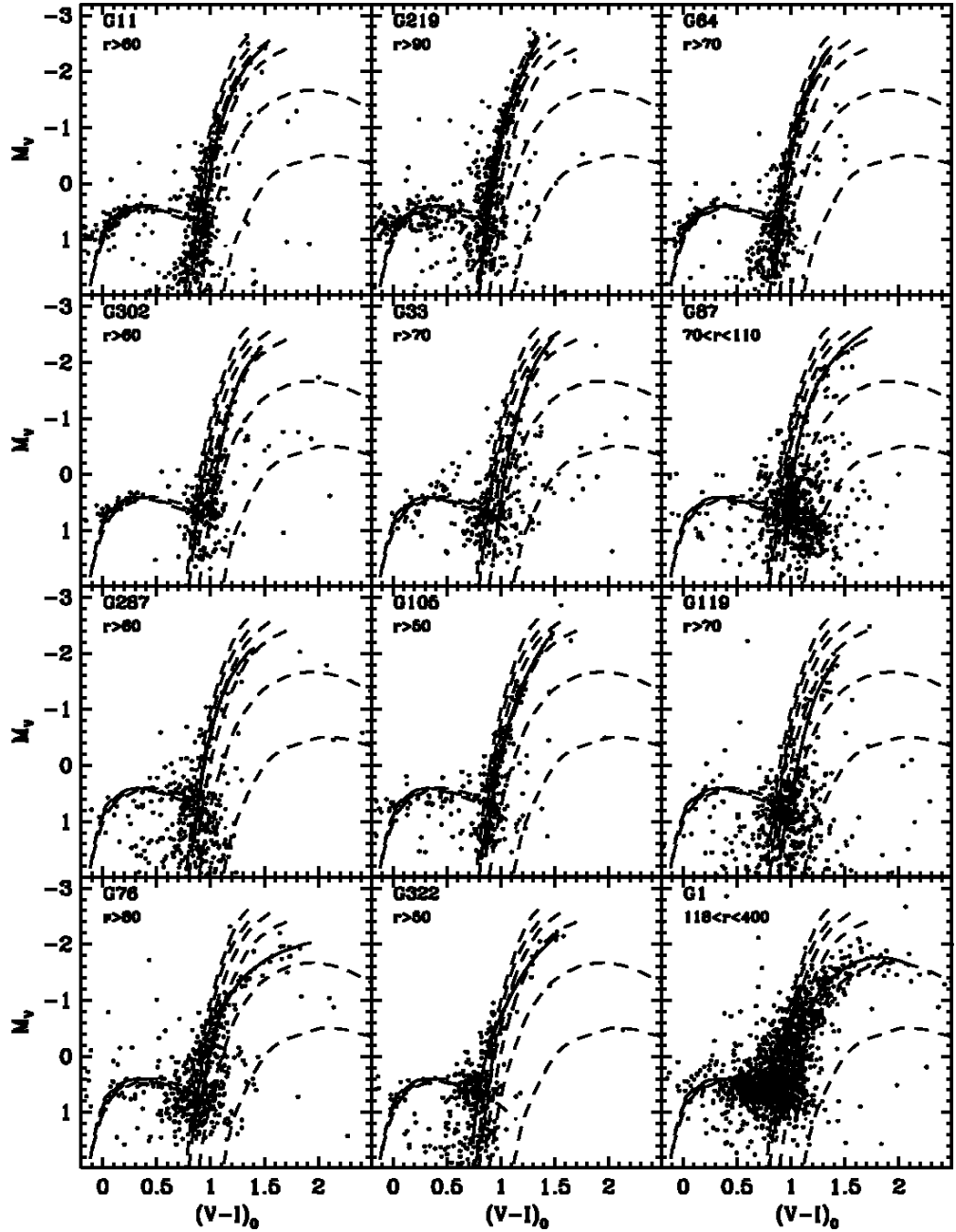}
\caption{RGB ridge lines of our GCs (shown as solid lines), plotted 
individually along with the grid of GGC RGB templates (shown as dashed 
lines). The (M$_V$/(V-I)$_0$ templates are for:  NGC6553, 47 Tuc, M5, M3, NGC6397 
and M15 (see Table \ref{tab:grid}). 
}
\label{fig:rgb1}
\end{figure}

\clearpage  

\begin{figure}
\plotone{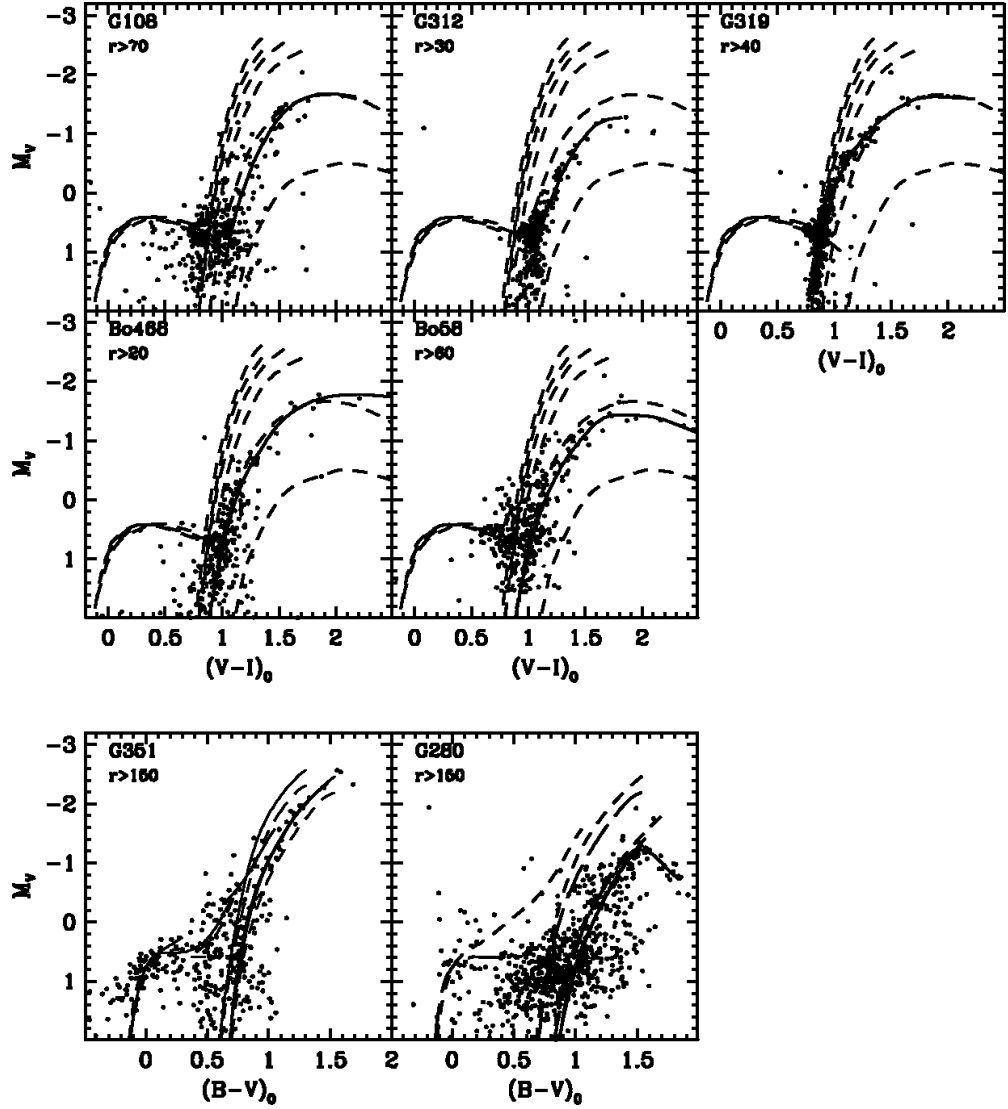}
\caption{RGB ridge lines of our GCs (shown as solid lines), plotted  
individually along with the grid of GGC RGB templates  (shown as dashed 
lines). The (M$_V$/(V-I)$_0$ templates are for:  NGC6553, 47 Tuc, M5, M3, 
NGC6397 and M15; the (M$_V$/(B-V)$_0$ templates are for: M15, NGC5824, M13, 
M5, 47 Tuc, NGC6356, and NGC6624 (see Table \ref{tab:grid}).  
}
\label{fig:rgb2}
\end{figure}

\clearpage

\begin{figure}
\plotone{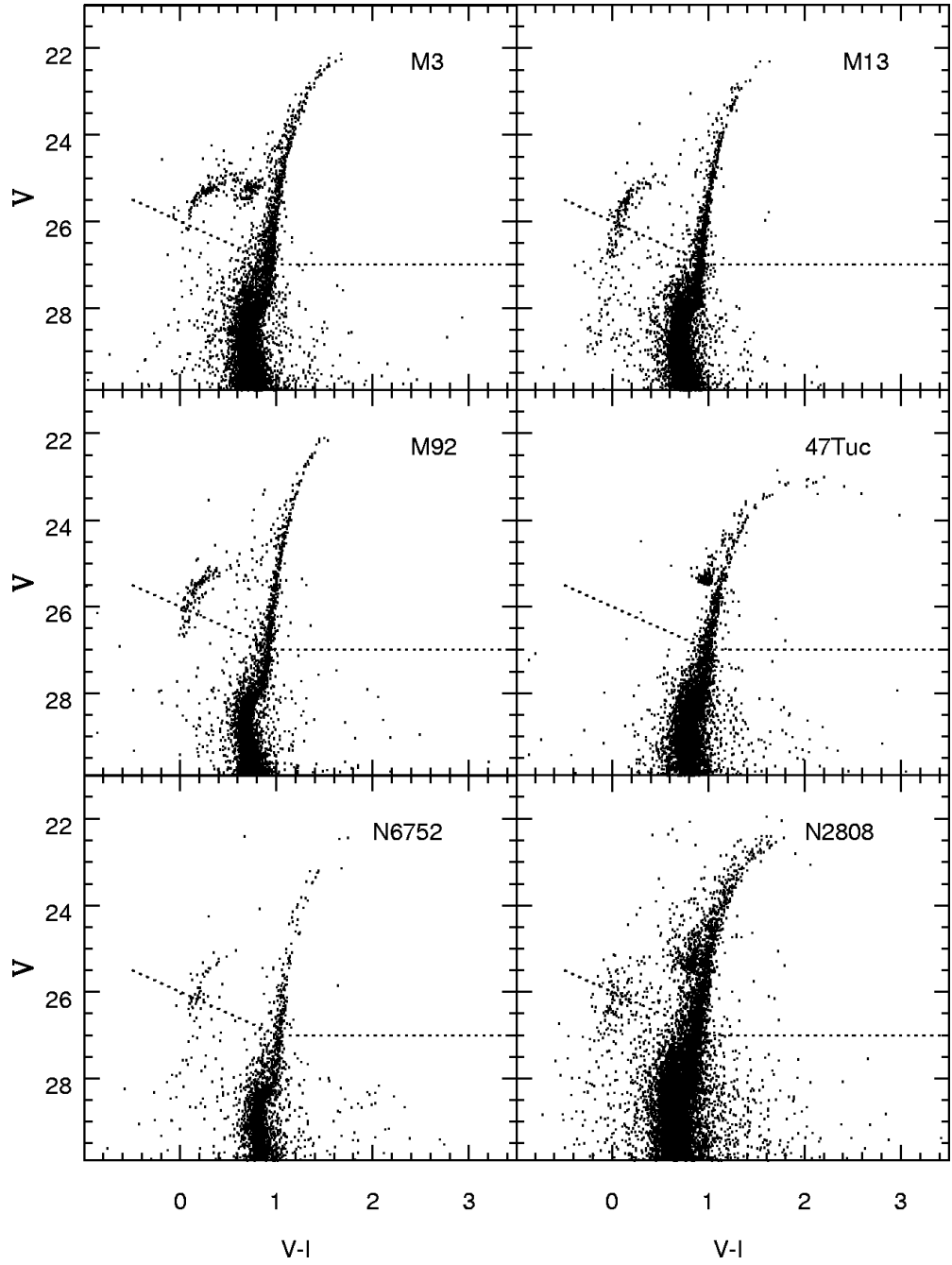}
\caption{Example  of  how  the  CMDs  of  6  among  the  best  studied  GGCs,  
namely  M3,  M13,  M92,  47  Tuc,  NGC6752  and  NGC2808,  would  appear  if  placed  at  
the  distance  of  M31  and  affected  by  the  photometric  cutoff  that  apply  
to  our  data  set.  
\label{fig:cutoff}}
\end{figure}

\clearpage

\begin{figure}
\plotone{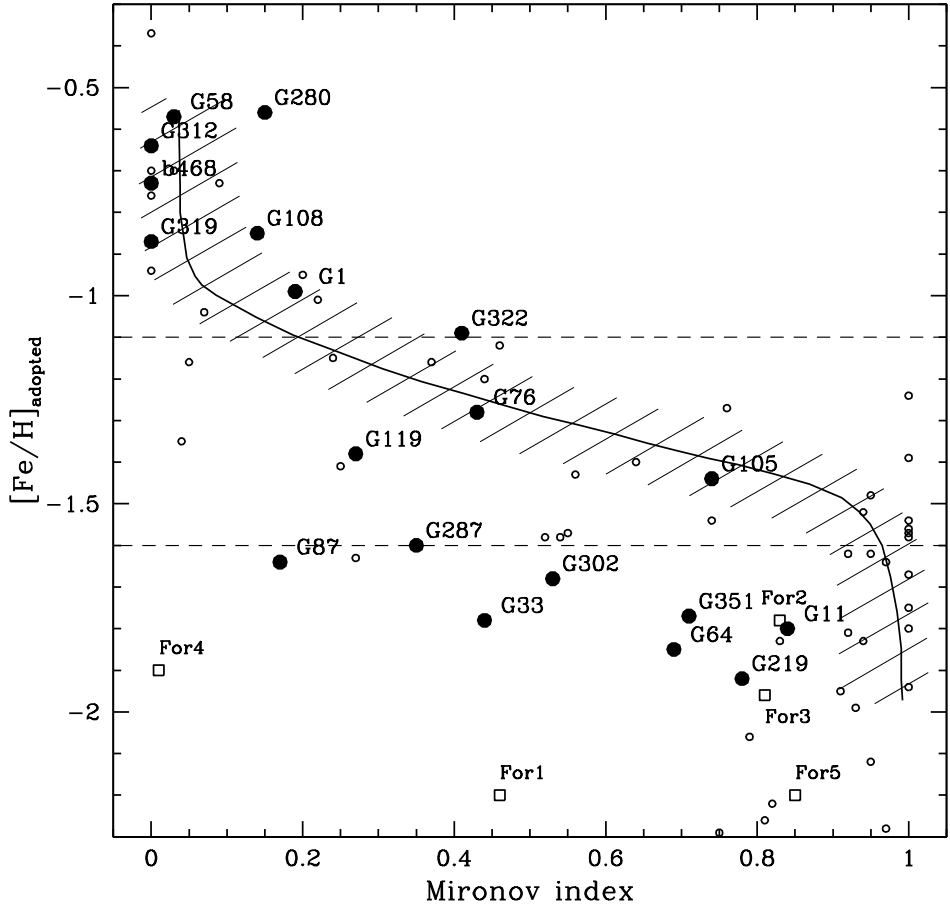}
\caption{HB  morphology, as described by the Mironov index B/(B+R),  
as a function of [Fe/H]. 
The  corresponding  behavior of the GCs in the MW (open circles) and in Fornax (open 
squares) 
is also shown for comparison (see Sect. 3.3.1).       
The 2nd parameter effect in M31 would appear to occur at lower metallicity  
than is the case for the Galaxy, consistent with expectations if the cluster  
system is younger.
}
\label{fig:hb}
\end{figure}

\clearpage  

\begin{figure}
\plotone{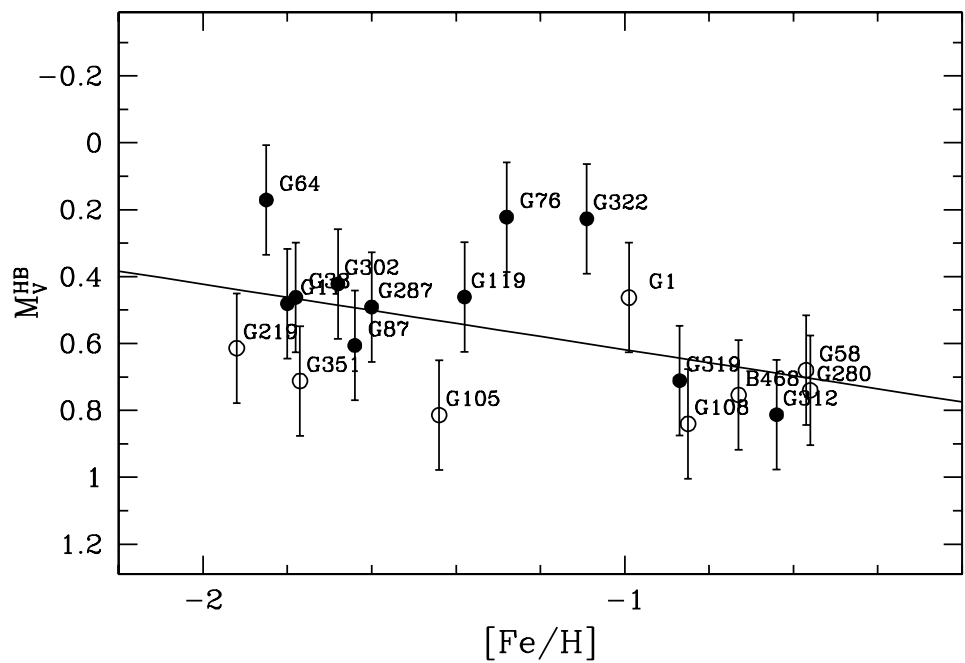}
\caption{Mean  M$_V$(HB) as a function of [Fe/H] for the present 11 GCs  
(filled  circles) and the 8 GCs that were studied in Paper I (open  circles).  
The line represents the best linear fit 
M$_V$(HB)=(0.20$\pm$0.09)[Fe/H]+(0.81$\pm$0.13), rms error of the fit 
$\sigma$=0.18. 
Some of the most discrepant clusters may be physically
located at a distance different from that of the main
M31 cluster population (see Sect. 3.3.2).  
\label{fig:mvfe}}
\end{figure}

\clearpage  


\begin{thebibliography}{}
\bibitem[Ajhar  et  al.(1996)]{ajh96}  Ajhar,  E.  A.,  Grillmair,  C.  J.,  Lauer,  T.  R.,    Baum,  W.  A.,  
      et  al.:  1996,    \aj,    111,  1110  
\bibitem[Ashman  et  al.(1998)]{ash98}  Ashman,  K.M.  \&  Zepf,  S.E.  1998,  in  {\it  Globular  Cluster  Systems},  
      (Cambridge:  Cambridge  University  Press)  
\bibitem[Barmby  \&  Huchra(2001)]{bhu01}  Barmby,  P.  \&  Huchra,  J.P.  2001,  \aj,  122,  2458  
\bibitem[Barmby  et  al.(2000)]{bhb00}  Barmby,  P.,  Huchra,  J.P.,  Brodie,  J.P.,  et  al.\  2000,  AJ,  119,  727
\bibitem[Barmby  et  al.(2001)]{bhb01}  Barmby,  P.,  Huchra,  J.P.  \&  Brodie,  J.P.  2001,  \aj,  121,  1482  
\bibitem[Barmby  et  al.(2002)]{bhh02}  Barmby,  P.,  Holland,  S.  \&  Huchra,  J.P.  2002,  \aj,  123,  1937  
\bibitem[Barmby(2003)]{bar03}  Barmby,  P.,  2003,  in  {\it  Extragalactic  Globular  
      Cluster  Systems},  ESO  Workshop,    ed.  M.  Kissler-Patig,  Springer-Verlag,  
      p.143 
\bibitem[Battistini et al.\ (1987)]{bat87} Battistini, P., B\`onoli, F., Braccesi, A., Federici, L., 
     Fusi Pecci, F., Marano, B. \& Borngen, F. 1987, \aaps, 67, 447 
\bibitem[Bekki  et  al.(2002)]{bek02}  Bekki,  K.,  Couch,  W.J.,  Drinkwater,  M.J.  \&  Gregg,  M.D.  2002,  
      \apj,  557,  L39  
\bibitem[Bellazzini  et  al.(1995)]{bel95}  Bellazzini,  M.,  Pasquali,  A.,  Federici,  L.,  Ferraro,  F.R.  \&  
    Fusi  Pecci,  F.  1995,  \apj,  439,  687  
\bibitem[Bellazzini  et  al.(1999)]{bel99a}  Bellazzini,  M.,  Ferraro,  F.R.  \&
            Buonanno,  R.  1999a,  \mnras,  304,  633  
\bibitem[Bellazzini  et  al.(1999)b]{bel99b}  Bellazzini,  M.,  Ferraro,  F.R.  \&
            Buonanno,  R.  1999b,  \mnras,  307,  619  
\bibitem[Bellazzini  et  al.(2001)]{bel01}  Bellazzini,  M.,  Ferraro,  F.R.  \&  Pancino,  E.  2001,  \apj,  556,  635  
\bibitem[Bellazzini  et  al.(2003)]{bel03}  Bellazzini,  M.,  Cacciari,  C.,  Federici,  L.,  Fusi  Pecci,  F.  \&  
      Rich,  M.  2003,  \aap,  405,  867
\bibitem[Benedict  et  al.(2002)]{ben02}  Benedict, G.F.  et  al.\  2002,  \aj,  123,  473  
\bibitem[Bonoli  et  al.(1987)]{bon87}  Bonoli,  F.,  Delpino,  F.E.,  Federici,  L.  \&  Fusi  Pecci,  F.,  1987,  
          \aap  185,  25
\bibitem[Brown  et  al.(2003)]{bro03}  Brown,  T.M.,  Ferguson,  H.C.,  Smith,  E.,  Kimble,  R.A.,  Sweigart,  A.V.,
      Renzini,  A.,  Rich,  R.M.  \&  VandenBerg,  D.A.  2003,  \apj,  592,  L17  
\bibitem[Brown  et  al.(2004a)]{bro04}  Brown,  T.M.,  Ferguson,  H.C.,  Smith,  E.,  Kimble,  R.A.,  Sweigart,  A.V.,
      Renzini,  A.,  Rich,  R.M.  \&  VandenBerg,  D.A.  2004a,  \aj,  127, 2738 
\bibitem[Brown  et  al.(2004b)]{bro04}  Brown,  T.M.,  Ferguson,  H.C.,  Smith,  E.,  Kimble,  R.A.,  Sweigart,  A.V.,
      Renzini,  A.,  Rich,  R.M.  \&  VandenBerg,  D.A.  2004b,  \apj, 613, L125 
\bibitem[Buonanno  et  al.(1983)]{buo83}  Buonanno,  R.,  Buscema,  G.,  Corsi,  C.E.
      \&  Iannicola  G.  1983,  \aap  126,  278
\bibitem[Buonanno  et  al.(1998)]{buo98}  Buonanno,  R., Corsi, C.E., Zinn, R., Fusi Pecci, 
     F., Hardy, E. \& Suntzeff, N.B. 1998, \apj, 501, L33
\bibitem[Burstein  \&  Heiles(1982)]{bhe82}  Burstein,  D.  \&  Heiles,  C.  1982,  \aj,  87,  1165  
\bibitem[Burstein  et  al.(1984)]{bfgk84}  Burstein,  D.,  Faber,  S.M.,  Gaskell,  C.M.  \&  Krumm,  N.  1984,  
    \apj,  287,  586  
\bibitem[Burstein et al.(2004)]{bfgk04} Burstein, D., Li, Y., Freeman, K.C., Norris, J.E. et al.\ 2004,  
    \apj, 614, 158 
\bibitem[Buzzoni  et  al.(1983)]{buz83}  Buzzoni, A.  et  al.\  1983,  \aap,  128,  94
\bibitem[Cacciari  \&  Clementini(2003)]{cc}  Cacciari,  C.  \&  Clementini,  G.  2003,  
      in  {\it  Stellar  Candles  for  the  Extragalactic  Distance  Scale},  Lect.  Notes  
      in  Phys.  Vol.  635,  eds.  D.  Alloin  and  W.  Gieren,  Springer-Verlag,  p.  105  
\bibitem[Carretta  \&  Gratton(1997)]{cg97}  Carretta  \&  Gratton  1997,  A\&AS,  121,  95 (CG97)
\bibitem[Choi  et  al.(2002)]{cgj}  Choi,  P.I.,  Guhathakurta,  P.  \&  Johnston,  K.V.  2002,  \aj,  124,  310    
\bibitem[Clementini  et  al.(2001)]{cle01}  Clementini,  G.,  Federici,  L.,  Corsi,  
      C.E.,  Cacciari,  C.,  Bellazzini,  M.  \&  Smith,  H.A.  2001,  \apj,  559,  L109  
\bibitem[Clementini  et  al.(2003)]{cle03}  Clementini,  G.,  Gratton,  R.G.,  
      Bragaglia,  A.,  Carretta,  E.,  Di  Fabrizio,  L.  \&  Maio,  M.  2003,  \aj,  125,  1309      
\bibitem[Corsi  et  al.(2000)]{cor00}  Corsi,  C.E.,  Rich,  M.R.,  Cacciari,  C.,  Federici,  L.  \&  Fusi  Pecci,  F.
      2000,  in  {\it  A  Decade  of  HST  Science},  Ed.s  M.  Livio,  K.  Noll  \&  M.  
      Stiavelli,  STScI  Symp.  Series  Vol.  14,  p.  28
\bibitem[Crampton  et  al.(1985)]{ccsc85}  Crampton,  D.,  Cowley,  A.P.,  Shade,  D.,
        \&  Chayer,  P.  1985,  \apj,  288,  494    
\bibitem[D'Antona \& Caloi (2004)]{da04} D'Antona, F. \& Caloi, V. 2004, \apj, 611, 871 
\bibitem[Di  Stefano  et  al.(2002)]{ds02}  Di  Stefano,  R.,  Kong,  A.K.H.,  Garcia,  M.R.,  Barmby,  P.,  Greiner,  J.,  
    Murray,  S.S.  \&  Primini,  F.A.  2002,  \apj,  570,  618  
\bibitem[Dolphin(2000)]{dol00}  Dolphin,  A.  E.  2000,  PASP,  112,  1383
\bibitem[Djorgovski  et  al.(1997)]{dj97}  Djorgovski,  S.G.,  Gal,  R.R.,  McCarthy,  J.K.,  Cohen,  J.G.,  de  
    Carvalho,  R.R.,  Meylan,  G.,  Bendinelli,  O.  \&  Parmeggiani,  G.    1997,  \apj,  
    474,  L19  
\bibitem[Djorgovski  et  al.(2003)]{dj03}  Djorgovski,S.G.,  C\^ot\'e,  P.,  Meylan,  G.,  Castro,  S.,  
   Federici,  L., et  al.\    2003  in    {\it  New  horizons  in  globular  cluster  astronomy}, 
   eds. G.Piotto, G. Meylan, S.G. Djorgovski and M. Riello,  A.S.P. Conf. Ser. Vol. 296, p. 479
\bibitem[Durrell et al. (2001)]{dur01} Durrell, P.R., Harris, W.E. \& Pritchet, C.J. 2001, \aj,  
     121, 2557
\bibitem[Ferguson  et  al.(2002)]{fer02}  Ferguson,  A.M.N.,  Irwin,  M.J.,  Ibata,  R.A.,  Lewis,  G.F.  \&  
      Tanvir,  N.R.  2002,  \aj,  124,  1452  
\bibitem[Ferraro  et  al.(1999)]{fe99}  Ferraro,  F.R.,  Messineo,  M.,  Fusi  Pecci,  F.,  De  Palo,  M.A.,  
      Straniero,  O.,  Chieffi,  A.  \&  Limongi,  M.,  1999,  \aj,    118,  1738  
\bibitem[Freedman et al. (2001)]{fre01} Freedman,  W.L. et al. 2001, \apj, 553, 47
\bibitem[Frogel  et  al.(1980)]{fro80}  Frogel,  J.A.,  Persson,  S.E.  \&  Cohen,  
      J.G.  1980,  \apj,  240,  785  
\bibitem[Fusi  Pecci  et  al.(1994)]{ffp94}  Fusi  Pecci,  F.,  Battistini,  P.,  Bendinelli,  O.,  
    Bonoli,  F.,  Cacciari,  C.,  Djorgovski,  S.G.,  Federici,  L.,  Ferraro,  F.R.,  
    Parmeggiani,  P.,  Weir,  N.,  \&  Zavatti,  F.  1994,  \aap,  284,  349
\bibitem[Fusi  Pecci  et  al.(1996)]{ffp96}  Fusi  Pecci,  F.,  Buonanno,  R.,  Cacciari,  C.,    
    Corsi,  C.  E.,  Djorgovski,  S.  G.,  Federici,  L.,  Ferraro,  F.  R.,  Parmeggiani,  
    G.,  Rich,  R.  M.  1996,    \aj,  112,  1461  (Paper  I)
\bibitem[Fusi  Pecci  \&  Bellazzini  (1997)]{ffp97}  Fusi  Pecci,  F.  \&  Bellazzini,  M.
     1997,  in  the  Third  Conf.  on  Faint  Blue  Stars    ed.  A.G.D.  Philip,  J.  Liebert,  \&  R.A.  Saffer
     (Schenectady:  L.  Davis  Press),  255
\bibitem[Fusi Pecci et al.(2004)]{ffp04} Fusi Pecci, F., Bellazzini, M., Buzzoni, A., De Simone, E., 
     Federici, L. \& Galleti, S. 2004, \aj, submitted     
\bibitem[Galleti  et  al.(2004)]{gaf04}  Galleti,  S.,  Federici,  L.,  Bellazzini,  M.,  Fusi  Pecci,  F.  \&  
    Macrina,  S.  2004,  \aap,  416,917  
\bibitem[Gratton  et  al.(2004)]{gra04}  Gratton,  R.G.,  Bragaglia,  A.,  
      Clementini,  G.,  Carretta,  E.,  Di  Fabrizio,  L.,  Maio,  M.  \&  Taribello,  E.  
      2004,  \aap,  421, 937
\bibitem[Guarnieri  et  al.(1998)]{gua98}  Guarnieri,  M.D.,  Ortolani,  S.,
    Montegriffo,  P.,  Renzini,  A.,  Barbuy,  B.,  Bica,  E.  \&  Moneti,  A.  1998,
  \aap,  331,  70
\bibitem[Harris(1996)]{ha96}  Harris,  W.E.  1996,  \aj,  112,  1487 
     (update 2003, http://physun.physics.mcmaster.ca/Globular.html) 
\bibitem[Harris  et  al.\  (1997)]{harris97}Harris,  W.E.,  Bell,  R.A.,  Vandenberg,  D.A.,  Bolte,  M.,
Stetson,  P.B.  et  al.\  1997,  \aj,  114,  1030
\bibitem[Hartwick(1968)]{har68}  Hartwick, F.D.A.,  1968,  \apj,  154,  475  
\bibitem[Holland (1998)]{hol98} Holland,  S. 1998, \aj, 115, 1916
\bibitem[Holland  et  al.(1997)]{hol97}  Holland,  S.,  Fahlman,  G.G.,  Richer,  H.B.  1997,    \aj,  114,  1488  
\bibitem[Huchra  et  al.(1991)]{huc91}  Huchra,  J.P.,  Brodie,  J.P.  \&  Kent,  S.M.  1991,  \apj,  370,  495  
\bibitem[Ibata  et  al.(1994)]{iba94}  Ibata,  R.A.,    Gilmore,  G.,  \&  Irwin,  M.J.  1994,  Nature,  370,  194  
\bibitem[Ibata  et  al.(2001)]{iba91}  Ibata,  R.A.,  Irwin,  M.J.,  Lewis,  G.F.,  Ferguson,  A.M.N.  \&  
      Tanvir,  N.R.  2001,  Nature,  412,  49    
\bibitem[Jablonka(1992)]{jab92}  Jablonka,  P.  ,  Alloin,  D.  \&  Bica.  E.  1992,  \aap,  260,  97
\bibitem[Jablonka  et  al.(2000)]{jab00}  Jablonka,  P.,  Courbin,  F.,  Meylan,  G.,  Sarajedini,  A.,  Bridges,  T.J.  
    \&  Magain,  P.    2000,  \aap,  359,  131   
\bibitem[Joshi et al. (2003)]{jos03} Joshi, Y.C., Pandey, A.K., Narasimha, D., 
      Sagar, R. \& Giraud-Héraud, Y. 2003, \aap,  402, 113
\bibitem[Lee  et  al.(1994)]{lee94}  Lee,  Y.-W.,  Demarque,  P.  \&  Zinn,  R.  1994,  
      \apj,  423,  248
\bibitem[Lupton (1989)]{lup89} Lupton, R. H. 1989, \aj, 97, 1350
\bibitem[McClure  \&  Racine(1969)]{mcr69}  McClure,  R.D.  \&  Racine,  R.  1969,  \aj,  74,  1000
\bibitem[McConnachie et al.  (2004)]{mcco04} McConnachie, A.W., Irwin, M.J., Ferguson, A.M.N., 
     Ibata, R.A., Lewis, G.F. \& Tanvir, N.  2004, \mnras, in press (astro-ph0410489)
\bibitem[McLaughlin(2000)]{mcl00}  McLaughlin,  D.E.  2000,  \apj,  539,  618
\bibitem[Meylan et al. (2001)]{mey01} Meylan, G., Sarajedini, A., Jablonka,  P., Djorgovski, S.G., Bridges,  T.J
     \& Rich, R.M. 2001, \aj, 112, 830  
\bibitem[Mighell  et  al.(1996)]{mig96}  Mighell,  K.J.,  Rich,  R.M.,  Shara,  M.  \&
            Fall.  S.M.  1996,  \aj,  111,  2314
\bibitem[Mironov(1972)]{mir72}  Mironov,  A.V.  1972,  Soviet  Astronomy,  16,  105
\bibitem[Moffat(1969)]{mof69}  Moffat,  A.F.J.  1969,  \aap,  3,  455  
\bibitem[Morrison  et  al.(2004)]{morr04}  Morrison,  H.L.,  Harding,  P.,  Perrett,  K.  \&  Hurley-Keller,  D.
          2004,  \apj,  603,  87  
\bibitem[Peterson  et  al.(2003)]{per03}  Peterson,  R.C.,  Carney,  B.W.,  Dorman,  B.,  Green,  E.M.,  Landsman,  W.,    
    Liebert,  J.,  O'Connell,  R.W.  \&  Rood,  R.T.  2003,  \apj,  588,  299
\bibitem[Perrett  et  al.(2002)]{per02}  Perrett,  K.M.,  Bridges,  T.J.,  Hanes,  D.A.,  Irwin,  M.J.,  Brodie,  
    J.P.,  Carter,  D.,  Huchra,  J.P.  \&  Watson,  F.G.  2002,  \aj,  123,  2490
\bibitem[Piotto  et  al.(2002)]{pio02}  Piotto,  G.,  King,  I.R.,  Djorgovski,  S.G.,
    Sosin,  C.,  Zoccali,  M.,  et  al.\  2002,  \aap,  391,  945  
\bibitem[Piotto et al.(2004)]{pio04} Piotto, G. et al.\ 2004, \apj, in press (astro-ph/0412016)
\bibitem[Rich (2004)]{ric04} Rich,  R.M. 2004, in {\it Origin and Evolution of the Elements}, ed. A. McWilliam 
     and M. Rauch, Carnegie Obs. Ap. Series Vol. 4, p. 258  (Cambridge: Cambridge Univ. Press)  
\bibitem[Rich  et  al.(1996)]{ric96a}  Rich,  R.M.,  Mighell,  K.J.,  Freedman,  W.,  \&  Neill,  J.D.  1996,  
  \aj,  111,  768    
\bibitem[Rich  et  al.(2001)]{ric01}  Rich,  R.M.,  Corsi,  C.E.,  Bellazzini,  M.,  Federici,    L.,  Cacciari,  C.  
    \&  Fusi  Pecci,  F.  2001,  in  {\it  Extragalactic  Star  Clusters},  eds.  E.  
    Grebel,  D.  Geisler  and  D.  Minniti,    IAU.  Symp.  207,  p.  140  
\bibitem[Rosenberg  et  al.(1999)]{ros99}  Rosenberg,  A.,  Saviane,  I.,  Piotto,  G.
      \&  Aparicio,  A.  1999,  \aj,  118,  2306  
\bibitem[Rosenberg  et  al.(2000a)]{ros00a}  Rosenberg,  A.,  Piotto,  G.,  Saviane,  I.
      \&  Aparicio,  A.  2000a,  \aaps,  144,  5
\bibitem[Rosenberg  et  al.(2000b)]{ros00b}  Rosenberg,  A.,  Aparicio,  A.,  Saviane,  I.
      \&  Piotto,  G.    2000b,  \aaps,  145,  451
\bibitem[Rosenberg  et  al.(2002)]{ros02}Rosenberg,  A.,  Aparicio,  A.,  Piotto,  G.,  \&  Saviane,  I.
     2002,  Ap\&SS  281,  125
\bibitem[Saito  \&  Iye(2000)]{sai00}  Saito,  Y.  \&  Iye,  M.  2000,  \apj,  535,  L95  
\bibitem[Sarajedini(1994)]{sar94}  Sarajedini,  A.,  1994,  AJ,  107,  618
\bibitem[Sargent et al.\ (1977)]{sar77} Sargent, W.L.W., Kowal, C.T., Hartwick, F.D.A. \& 
     van den Bergh, S. 1977, \aj, 82, 947 
\bibitem[Saviane  et  al.(2000)]{sav00}  Saviane,  I.,  Rosenberg,  A.,  Piotto,  G.  \&  Aparicio,  A.    2000,  
        \aap,  355,  966
\bibitem[Schlegel  et  al.(1998)]{sch98}  Schlegel,  D.J.,  Finkbeiner,  D.P.  \&
      Davis,  M.  1998,  \aj,  500,  525  
\bibitem[Smith  et  al.\  1996]{smith96}Smith,  E.O.,  Neill,  J.D.,  Mighell,  K.J.,  \&  Rich,  R.M.  1996,  \aj,  111,  1596
\bibitem[Stanek \& Garnavich (1998)]{sg98} Stanek, K.Z. \& Garnavich, P.M. 1998, \apj, 503, L131 
\bibitem[Staneva et al.\ (1996)]{sta96} Staneva, A., Spassova, N. \& Golev, V. 1996, \aaps, 
     116, 447
\bibitem[Stephens  et  al.(2001)]{ste01}  Stephens,  A.W.,  Frogel,  J.A.,  Freedman,  W.,  Gallart,  C.,  Jablonka,    P.,  Ortolani,  S.,  Renzini,  A.,  Rich,  R.M.,  Davies,  R.  2001,  \aj,  121,  2597
\bibitem[van  den  Bergh(1969)]{vdb69}  van  den  Bergh,  S.  1969,  ApJS  19,145
\bibitem[van  den  Bergh  (2000)]{vdb}  van  den  Bergh,  S.  2000,  PASP,  112,  932
\bibitem[van  den  Bergh  (2003)]{vdb03}  van  den  Bergh,  S.  2003,  in {\it The Local Group as an Astrophysical 
     Laboratory}, (Cambridge: Cambridge Univ. Press), in press (astro-ph/0305042)
\bibitem[van  Speybroeck  et  al.(1979)]{vsp79}  van  Speybroeck,  L.  et  al.\  1979,  \apj,  234,  L45  
\bibitem[Vetesnik(1962)]{vet62}Vetesnik,  M.  1962,  Bull.Astron.Inst.Czechoslovakia.,13,180
\bibitem[Zinn  \&  West(1984)]{zw84}  Zinn,  R.J.  \&  West,  M.J.  1984,  \apjs,  55,  45 (ZW84) 
\bibitem[Zinn(1985)]{z85}  Zinn,  R.J.  1985,  \apj,  293,  424
\bibitem[Trudolyubov  \&  Priedhorsky(2004)]{trp04}  Trudolyubov,  S.  \&  
Priedhorsky, W. 2004,  \apj, in press (astro-ph/0402619)
\end{thebibliography}
\end{document}